\documentclass[journal]{IEEEtran}
\usepackage{amsfonts}
\usepackage{amssymb}
\usepackage[cmex10]{amsmath}
\usepackage{graphicx}
\usepackage{subfigure}
\usepackage{cite}
\usepackage{tabularx,booktabs}
\usepackage{amsthm}
\usepackage{booktabs}
\usepackage{multirow}
\usepackage{setspace}
\usepackage{fancyhdr}
\usepackage{algorithmic}
 \usepackage{enumerate}
 \usepackage{color}
\usepackage{textcomp}

\IEEEoverridecommandlockouts
\newcommand{\RNum}[1]{\uppercase\expandafter{\romannumeral #1\relax}}
\newcommand{\be}{\begin{equation}}
\newcommand{\ee}{\end{equation}}

\allowdisplaybreaks
\ifCLASSINFOpdf
\else
\fi
 \usepackage{nomencl}
 \makenomenclature

\begin{document}

\title{Supervised Learning Based Online Tracking Filters: An XGBoost Implementation}

\author{Wei~Yi,
        Jie~Deng
\thanks{W. Yi, J. Deng are with the School of Information and Communication Engineering, the University of Electronic Science and Technology of China, Chengdu 611731, China. (email: kussoyi@gmail.com; jiedeng\_sc@outlook.com)}}

\maketitle

\begin{abstract}
The target state filter is an important module in the traditional target tracking framework. In order to get satisfactory tracking results, traditional Bayesian methods usually need accurate motion models, which require the complicated prior information and parameter estimation. Therefore, the modeling process has a key impact on traditional Bayesian filters for target tracking. However, when encountering unknown prior information or the complicated environment, traditional Bayesian filters have the limitation of greatly reduced accuracy. In this paper, we propose a supervised learning based online tracking filter(SLF). First, a complete tracking filter framework based on supervised learning is established, which is directly based on data-driven and establishes the mapping relationship between data. In other words, the proposed filter does not require the prior information about target dynamics and clutter distribution. Then, an implementation based on eXtreme Gradient Boosting (XGBoost) is provided, which proves the portability and applicability of the SLF framework. Meanwhile, the proposed framework will encourage other researchers to continue to expand the field of combining traditional filters with supervised learning. Finally, numerical simulation experiments prove the effectiveness of the proposed filter.
\end{abstract}

\begin{IEEEkeywords}
Tracking filter, hidden modeling, XGBoost, data-driven.
\end{IEEEkeywords}

\IEEEpeerreviewmaketitle


\section{Introduction}\label{sec:intro}

\IEEEPARstart{T}{he} filtering is a data processing technology that removes noise and restores the real data. It has been widely used in target tracking, computer vision, and defense guidance. For the estimation problem in time series, the filter uses the historical measurements to estimate the target state. In addition, the Bayesian filter is a classic filtering framework \cite{MS2002}, which is widely used in the target tracking. Nowadays, a series of target tracking filters have been developed based on this framework. For example, the Kalman filter (KF) \cite{Kalman1960}, the extended Kalman filter (EKF) \cite{Kalman1961}, the unscented Kalman filter (UKF) \cite{SJ2004}, and the particle filter (PF) \cite{MS2002}.

Generally, the traditional Bayesian filter require two models in target tracking fields: 1. the motion model (a model describing the evolution of the state with time); 2. the measurement model (a model relating the noisy measurements to the state). Therefore, when using the Bayesian filter, it is necessary to model the above two basic models and estimate the basic parameters of the models. For example, the process noise intensity and the measurement noise covariance. In this paper, we call the process to build the model and estimate prior parameters as ``visible modeling''. In order to get accurate tracking results, ``visible modeling'' requires that the model must reasonably match the actual system as much as possible, which is also a limitation of traditional Bayesian filters. In order to address this problem, some sophisticated models or adaptive multiple models have been proposed. The motion models of various targets were considered in \cite{X2003} and \cite{X2010}. Li~\emph{et~al}.~\cite{X2005} focused on the multi-model approach.  Jwo~\emph{et~al}.~\cite{Jwo2004} proposed training a multi-layer neural network to identify the measurement noise covariance matrix. Meanwhile, adaptive methods were also widely used. An adaptive extended Kalman filter was used to estimate the process noise of the target in \cite{Lee2010}. Tripathi~\emph{et~al}.~\cite{Tripathi2016} proposed an adaptive filter for the unknown noise. However, the implementation of the above method is at the cost of increasing the complexity and ``volume'' of the model, which also leads us to consider a new solution way.

Supervised learning is the machine learning \cite{Hastie2009}, task of inferring a function from labeled training data \cite{LeCun2015}. The training data consist of a set of training examples. Each example is a pair consisting of an input object and a desired output object. A supervised learning algorithm analyzes the training data and produces an inferred function, which can be used for mapping new examples. There are many widely used supervised learning
algorithms, such as the Logistic Regression(LR) \cite{David2000}, the Support Vector Machine (SVM) \cite{Cortes1995}, the Neural Network(NN) \cite{Simon1998} and the Decision Trees(DT) \cite{JR1986}. Nowadays, they have been widely used in many fields (e.g., text categorization \cite{M2009}, speech recognition \cite{Ldeng2013} and image processing \cite{Carneiro2007}).

Regarding the application of supervised learning in target tracking, most of the current research focuses on target tracking in the image and visual fields \cite{Li2016},  \cite{Milan2017}, \cite{Son2015}. In addition, some researches that apply supervised learning to filtering mostly choose neural networks as auxiliary algorithms. For example, Chin~\emph{et~al}.~\cite{Chin1994} first used NN to learn the residuals between state estimates and predicted values. For the lack of measurement information when the GPS signal is interrupted, the trained network is used to compensate for the EKF proposed in \cite{Haidong2013}. Recently, a filter for mapping target values directly to estimated states based on random forest (RF) was proposed in \cite{Thormann2017}. Then, Zhai~\emph{et~al}.~\cite{Zhai2019}  based on \cite{Thormann2017}, using the XGBoost instead of the RF to improve the accuracy of simulation results for the same problem. Meanwhile, Gao~\emph{et~al}.~\cite{Gao2019} considered using long short term memory (LSTM) to address the filtering problem of target tracking, which has a good estimation effect. Although supervised learning has achieved good results in image and video tracking, few scholars studied in the point-filtering. A few studies have been limited to the neural network's assistance or the residual training \cite{Chin1994}, \cite{Haidong2013}. What's more, none of the above-mentioned work \cite{Thormann2017} -- \cite{Gao2019} extracted the essential information in the filtering problem, and did not consider the sample sparseness problem.

In this paper, we propose a supervised learning based online tracking filter (SLF), which is implemented by the XGBoost algorithm. The proposed filter use supervised learning to analyze the data and build internal mapping relationships. After training based on the existing data, it can estimate new measurements and implement filtering. The proposed filter can avoid modeling the motion system by building a data mapping relationship and get rid of the limitations of traditional model-based filters. Therefore, this method is called a ``hidden modeling'' (i.e., it does not require prior model about target dynamics and clutter distributions). The main contributions of the paper are given as follows:

\begin{itemize}
        \item[1)] \emph{A specific framework for SLF is established}: We start from the processing of the underlying training data, propose a complete set of theoretical support and algorithm application frameworks. In the field of combining supervised learning with traditional filters, a new solution framework and theoretical method are successfully constructed.
        \item[2)] \emph{The extraction of essential motion information is solved}: For filters based on supervised learning, we extract the target motion information from three aspects of the time, the space and the angle, which is helpful for training of supervised learning algorithms. This is also where \cite{Thormann2017} -- \cite{Gao2019} did not consider.
        \item[3)] \emph{Take XGBoost as an example to implement the above framework}: The hypothetical function parameters and loss functions of the XGBoost algorithm are discussed in detail. Thus, the SLF framwork based on the XGBoost implementation is given, so as to propose an idea to further transplant new supervised learning into traditional filtering and tracking methods.
\end{itemize}

The rest of the paper is organized in the following manner. The traditional Bayesian estimation based tracking filter is detailed in Section~\ref{Sec:II} which also includes the limitation of model-based filters. A supervised learning based online tracking filter (SLF) is developed in Section~\ref{Sec: III}. In Section~\ref{Sec: XGBoost}, use XGBoost as a specific implementation of the SLF framework. The simulation results are presented in Section~\ref{Sec: simulation}. Conclusions are given in Section~\ref{Sec: conclusion}.

\section{Bayesian Estimation Based Tracking Filters }\label{Sec:II}

Consider a system, whose state space equation and measurement equation \cite{MS2002} are
\begin{align}
{{\bf{x}}_k} &= {f_k}\left( {{{\bf{x}}_{k - 1}},{{\bf{w}}_{k - 1}}} \right),\\
{{\bf{z}}_k} &= {h_k}\left( {{{\bf{x}}_k},{{\bf{v}}_k}} \right),
\end{align}
where ${f_k}$ is the state transition function, $\left\{ {{{\bf{w}}_{k - 1}},k \in \mathbb{N}} \right\}$ is an independent identical distribution (i.i.d) process noise, ${h_k}$ is the measurement function, $\left\{ {{{\bf{v}}_k},k \in \mathbb{N}} \right\}$ is an i.i.d measurement noise, $\mathbb{N}$ is the set of natural numbers, ${{\bf{x}}_k}$ is the target state at time $k$, ${{\bf{z}}_k}$ is the measurement at time $k$.

The above (1) and (2) can also be described by transition probability as $p\left( {{{\bf{x}}_k}|{{\bf{x}}_{k - 1}}} \right)$ and $p\left( {{{\bf{z}}_k}|{{\bf{x}}_k}} \right)$.

The purpose of filter is to recursively estimate the state ${{\bf{x}}_k}$ from the measurement ${{\bf{z}}_{1:k}}$, where ${{\bf{z}}_{1:k}} = \left\{ {{{\bf{z}}_i},i = 1,...,k} \right\}$ is the set of all available measurements up to time $k$. More specifically, the essence of Bayesian filter is to use the system model to predict the state's prior probability density and the latest measurements are updated to get the posterior probability density function.

Herein, the measurement ${{\bf{z}}_{1:k}}$ is used to recursively calculate the credibility when the state ${{\bf{x}}_k}$ takes different values to obtain the optimal estimate. Therefore, a probability density function $p\left( {{{\bf{x}}_k}|{{\bf{z}}_{1:k}}} \right)$ is constructed. Assuming that the $p\left( {{{\bf{x}}_{k-1}}|{{\bf{z}}_{1:k-1}}} \right)$ is obtained at time $k - 1$, the system model (1) are used to obtain the prior probability distribution of the state at time $k$ as follows
\begin{align}\label{eq3}
p\left( {{{\bf{x}}_k}|{{\bf{z}}_{1:k - 1}}} \right) = \int {p\left( {{{\bf{x}}_k}|{{\bf{x}}_{k - 1}}} \right)} p\left( {{{\bf{x}}_{k - 1}}|{{\bf{z}}_{1:k - 1}}} \right)d{{\bf{x}}_{k - 1}},
\end{align}
where $p\left( {{{\bf{x}}_k}|{{\bf{x}}_{k - 1}}} \right) = p\left( {{{\bf{x}}_k}|{{\bf{x}}_{k - 1}},{{\bf{z}}_{1:k - 1}}} \right)$.

A new measurement ${{\bf{z}}_k}$ can be obtained at time $k$. Based on the ``Bayesian approach'', the measurement model (2) can be used to update the prior probability distribution to obtain the state estimation as follows
\begin{align}\label{eq4}
p\left( {{{\bf{x}}_k}|{{\bf{z}}_{1:k}}} \right) = \frac{{p\left( {{{\bf{z}}_k}|{{\bf{x}}_k}} \right)p\left( {{{\bf{x}}_k}|{{\bf{z}}_{1:k - 1}}} \right)}}{{p\left( {{{\bf{z}}_k}|{{\bf{z}}_{1:k - 1}}} \right)}},
\end{align}
where
\begin{align}\label{eq5}
p\left( {{{\bf{z}}_k}|{{\bf{z}}_{1:k - 1}}} \right) = \int {p\left( {{{\bf{z}}_k}|{{\bf{x}}_k}} \right)p\left( {{{\bf{x}}_k}|{{\bf{z}}_{1:k - 1}}} \right)} d{{\bf{x}}_k}.
\end{align}

In summary, \eqref{eq3} and \eqref{eq4} are the two basic steps of Bayesian filter, the recursive calculation of \eqref{eq3} and \eqref{eq4} constitutes the optimal Bayesian estimation. Meanwhile, according to the minimum mean square (MMSE) criterion, the state with the maximum posterior probability density is used as the optimal estimation as follows
\begin{align}\label{eq6}
{{\bf{\hat x}}_k} = E\left[ {{{\bf{x}}_k}\left| {{{\bf{z}}_{1:k}}} \right.} \right] = \int {{{\bf{x}}_k}p\left( {{{\bf{x}}_k}\left| {{{\bf{z}}_{1:k}}} \right.} \right)} d{{\bf{x}}_k}.
\end{align}

However, this recursive propagation of posterior density is only a conceptual solution. Therefore, it is only possible to obtain an analytical calculation method based on the assumption of a specific distribution, such as a Gaussian distribution. The Kalman filter is the analytical calculation method of the Bayesian filter under the Gaussian distribution.

In addition, it can be clearly seen from the above theory that the Bayesian filter is a model based filter method. The Bayesian filter requires a reasonable matching of a motion model. Therefore, this traditional Bayesian filter has some limitations for unknown motion models or complicated noise environments.

\section{Supervised Learning Based Online Tracking Filters (SLF)}\label{Sec: III}
\subsection{Supervised Learning}\label{Sec: analysis-A}

Supervised learning is the machine learning task of learning a function that maps an input to an output based on example input-output pairs \cite{Pete2009}. Supervised learning includes the following elements: the input feature, the output variable, the hypothesis function and the loss function.

Suppose the input feature is defined as ${\bf{a}}$ and the output variable is ${\bf{b}}$. This output variable is the corresponding fact output in the training set. Meanwhile, the $i$-th input feature vector of ${\bf{a}}$ is written as
\begin{align}\label{eq7}
{{\bf{a}}^{(i)}} = {\left( {a_1^{(i)},a_2^{(i)}, \ldots ,a_j^{(i)}, \ldots ,a_n^{(i)}} \right)^{\top}},
\end{align}
where ``$^{\top}$'' denotes the matrix transpose, $a_j^{(i)}$ represents the $j$-th feature of ${{\bf{a}}^{(i)}}$, $i$ is the $i$-th sample, $\left( {{\bf{a,b}}} \right)$ is a sample.

Use the hypothetical function $h:{\bf{a}} \to {\bf{b}}$ to construct a ``hidden model'' between $\bf{a}$ and $\bf{b}$. More specifically, $h\left( {{\bf{a}},\phi } \right)$ is an estimator corresponding to the output variable ${\bf{b}}$, where $\phi$ denotes the parameter in $h$. Introduce the loss function to evaluate the quality of the parameter $\phi$ (i.e., find a suitable set of parameters so that the loss function is minimized). Thus, the loss function is defined as
\begin{align}\label{eq8}
J\left( \phi  \right) = \sum\limits_i {l\left( {{b_i},h\left( {{\bf{a,}}\phi } \right)} \right)}  + \lambda \Omega \left( \phi  \right),
\end{align}
where ${b_i}$ is the real output corresponding to the $i$-th input feature, $l\left(  \cdot  \right)$ is the training error term (e.g., square error function, the Logistic loss function \cite{LeCun2015}), $\Omega \left( \phi  \right)$ is the regularization factor, which denotes the complexity of the model and prevents the over-fitting, $\lambda $ controls the trade-off between data fitting error and model fitting error \cite{Li2019}.

Therefore, the advantage of supervised learning is that it can directly construct the mapping relationship between data based on data-driven. Then, it generates reasonable predictions for new inputs using the mapping.

Based on supervised learning' advantages, it is widely used in high-latitude data classification and forecasting fields, such as the big data processing \cite{He2016}, the image processing \cite{Carneiro2007} and the demand forecasting \cite{Sharma2011}. Take the house prices prediction as an example, some features related to house prices (e.g., the house area, the geographical location and the number of rooms) can be used as input features, the corresponding house prices are used as output variables. A high-dimensional mapping relationship between features and corresponding house prices is established by training algorithms. Therefore, we can predict new house prices for some new samples, which is more accurate than traditional complex models and multivariate functions \cite{Phan2018}.

\subsection{SLF Whole Ideas}\label{Sec: analysis-B}
On the one hand, because supervised learning based on data-driven, it can get rid of the limitations of  model-based filters and does not require model matching. On the other hand, because the traditional filtering is also a method of finding a generalized function based on historical data and outputting the predicted state. Meanwhile, the regression technique of supervised learning has a good effect on the prediction of continuous response. Therefore, we have reason to combine the filtering problem with supervised learning. Based on these, we propose a basic idea of supervised learning based online tracking filter (SLF).

For the filtering problem of target tracking, we use the sensor measurement ${\bf{z}}$ as the input feature (i.e., ${\bf{z}}$ is $\bf{a}$ in Section~\ref{Sec: analysis-A}) after preprocessing, the error ${\bf{r}}$ between the true state and the measurement as the output variable for increase the prediction generalization ability. Further, we train a hypothesis function $h\left( {{\bf{z}},\phi } \right)$ to build a ``hidden model'' with the optimal parameter $\hat \phi$. Subsequently, the trained hypothesis function is used to estimate the new measurement ${{\bf{z}}^{\rm{n}}}$, where ``$\rm{n}$'' stands for new. Based on the trained hypothesis function, we can get the estimated value ${\bf{\hat r}}$. Finally, we get the state estimation ${\bf{\hat x}}$ after some processing

SLF is mainly based on a supervised learning framework, which can be roughly divided into the following three steps:

\textbf{Step 1} -- \emph{SLF data preprocessing phase:} If the measurement is directly used as the training input feature, each sample will have only one feature and the algorithm cannot train it. Because in the supervised learning framework, each sample needs to have some features to represent the ``information and characteristics'' of it. Therefore, to address the filtering problem of target tracking, it is necessary to artificially preprocess the data and design input features for each point (sample) to express its information.

\textbf{Step 2} -- \emph{SLF training phase:} Herein, we construct the ``hidden model'' (i.e., the hypothesis function with parameters). Then, we train it using the loss function corresponding to the training algorithm (e.g., XGBoost is a training algorithm based on regression trees). Thus, the training algorithm continuously searches for an optimal parameter according to the loss function minimum principle.

\textbf{Step 3} -- \emph{SLF estimation and application phase:} The new measurement is input into the trained hypothesis function to get the estimated value. And the state estimation is obtained by inverse transformation in Section~\ref{Sec: analysis-E}.

\subsection{SLF Data Preprocessing Phase}\label{Sec: analysis-C}
Let ${\bf{z}}_k^j$ denotes the measurement of the $j$-th track at time $k$, each measurement is composed of the x and y coordinate value. ${\bf{z}}_{1:k}^j = \left\{ {{\bf{z}}_i^j,i = 1, \ldots ,k} \right\}$ is the measurement sequence of the target. Let ${\bf{x}}_k^j$ denotes the true state of the $j$-th track at time $k$. Use ${\bf{\hat x}}_k^j$ to denote the state estimation and ${\bf{\hat x}}_{k + 1\left| k \right.}^j$ to denote the state prediction. And ${\bf{r}}_k^j = {\bf{x}}_k^j - {\bf{z}}_k^j$ represents the error between the true state and the measurement.

\subsubsection{Sample Sparseness in the Filtering of Target Tracking}\label{C.1}

For the filtering of target tracking, if it is to be combined with supervised learning, the problem of sample sparseness needs to be considered. The core lies in three aspects: the time, the space and the angle.

\begin{figure}[t]
\centering
\includegraphics[width=8cm]{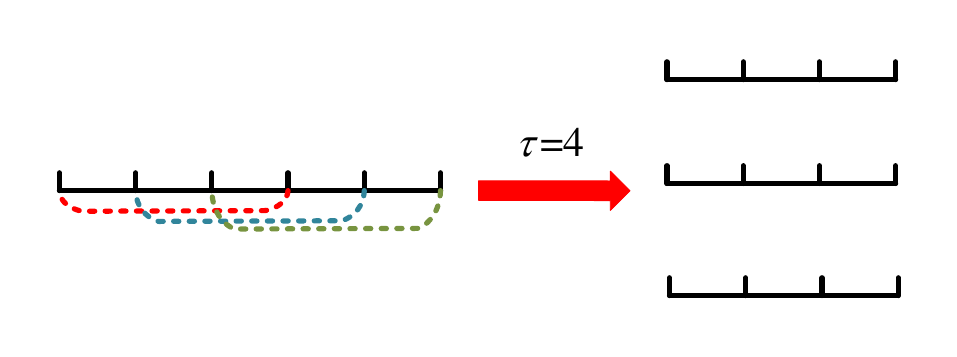}
\caption{The sliding window method, solve the problem of different lengths (different time) of track.}
\label{fig:sliding}
\end{figure}

\textbf{Step 1} -- \emph{Different track lengths (different time)}: The measurement track obtained by the sensor may have different lengths. \textbf{Confrontation scheme:} Use the ``sliding window'' method below, fixed $\tau$ measurements are intercepted from the measurement sequence of each track ($\tau$ represents the length of the sliding window) as shown in Fig. \ref{fig:sliding}.

\textbf{Step 2} -- \emph{Different track positions (different space)}: Due to the randomness of the initial position, the basic position of each track is different. \textbf{Confrontation scheme:} The ``relative measurement'' method below is adopted to extract the relative motion information. Thus, all input feature elements only store the relative displacement.

\begin{figure}[htbp]
\centering
{\includegraphics[width=8cm]{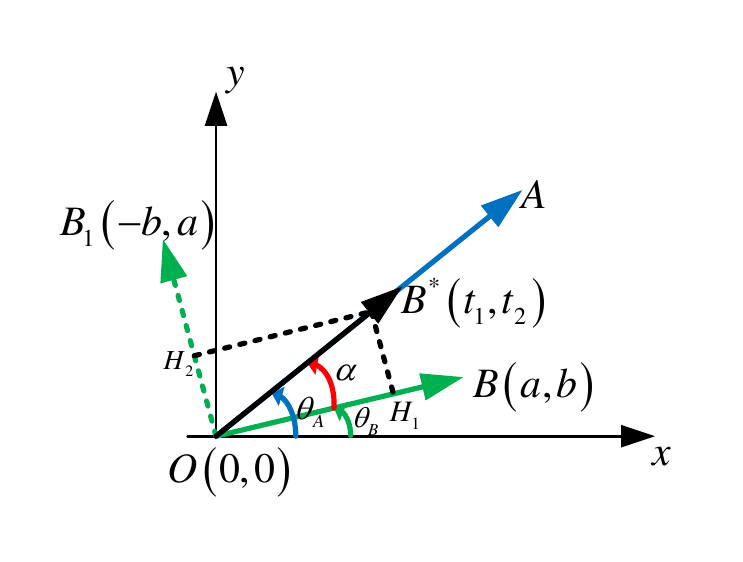}}
\caption{The rotation mapping method, mainly for the special case where the training sample track and the test sample track are in different initial directions, but have similar motion characteristics.}
\label{fig:rotation}
\end{figure}

\textbf{Step 3} -- \emph{Different track directions (different angles)}: Due to the different initial state, the direction of the track is different. However, the information in supervised learning only needs to represent the basic characteristics of motion. Therefore, the training samples with the same motion characteristics but different initial directions need to be considered. \textbf{Confrontation scheme:} Use the ``rotation mapping'' method below. Take the rotation of a two-dimensional vector as an example. As shown in Fig. \ref{fig:rotation}, consider rotating $\overrightarrow {OB}$ to the $\overrightarrow {OA}$ direction. The rotation angle is $\alpha$, $\overrightarrow {OB}$ becomes $\overrightarrow {O{B^{\rm{*}}}}$ after rotation. The angle of $\overrightarrow {O{B^{\rm{*}}}}$ is $ {\theta _B}$, the angle of $\overrightarrow {OA}$  is ${\theta _A}$. Suppose ${\theta _B} < {\theta _A}$, the rotation is counterclockwise and $\alpha  = {\theta _A} - {\theta _B}$ is set to positive angle.

Set $\overrightarrow {O{B^{\rm{*}}}}  = \left( {{t_1},{t_2}} \right)$ after rotation, and
\begin{align}
\label{eq9}{t_1} &= a\cos \alpha  - b\sin \alpha, \\
\label{eq10}{t_2} &= b\cos \alpha  + a\sin \alpha.
\end{align}

When ${\theta _B} > {\theta _A}$, $\alpha$ is set to a negative angle (clockwise rotation), the coordinate formulas after rotation are still \eqref{eq9} and \eqref{eq10}. In addition, we can also get the ``inverse rotation mapping'' that return from $\overrightarrow {O{B^{\rm{*}}}}$ to the original $\overrightarrow {OB}$, the ``inverse rotation mapping'' is given by
\begin{align}
\label{eq11}a &= {t_1}\cos \alpha  + {t_2}\sin \alpha, \\
\label{eq12}b &= {t_2}\cos \alpha  - {t_1}\sin \alpha.
\end{align}

\begin{figure}[htbp]
\centering
{\includegraphics[width=8cm]{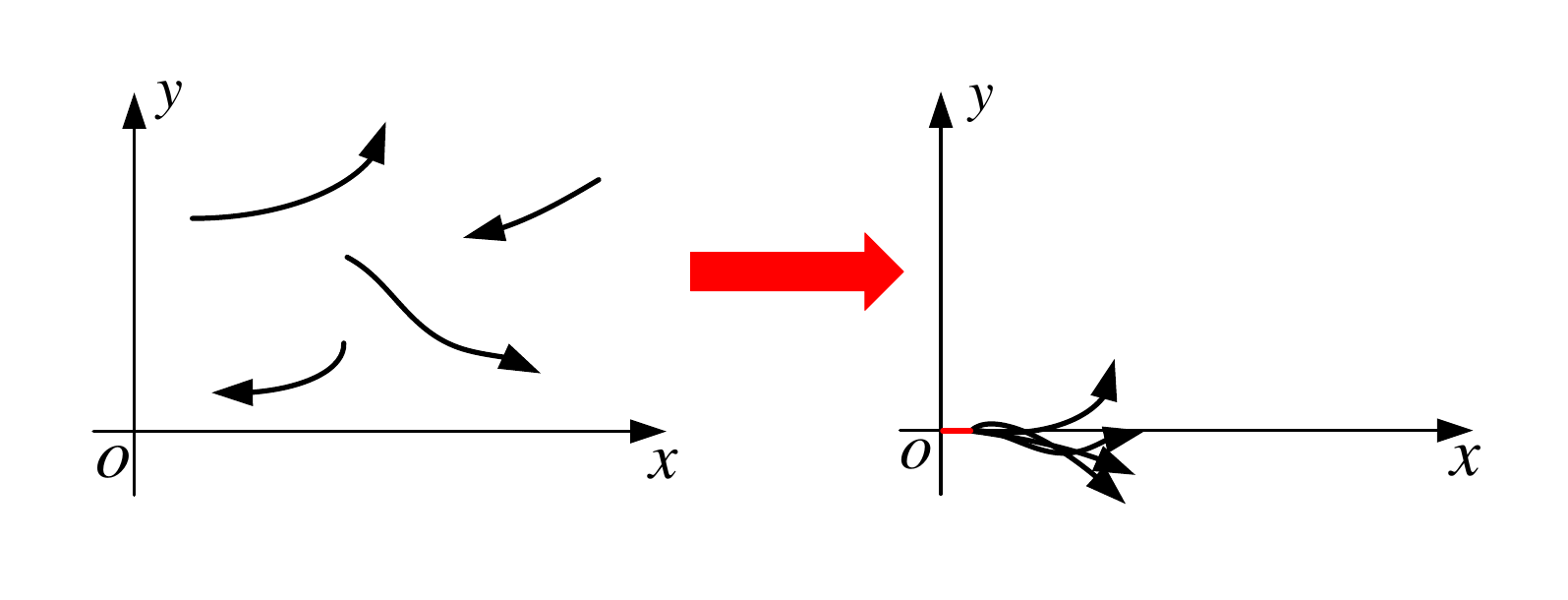}}
\caption{The processing result of the sample sparseness confrontation, mainly to better extract the essential motion information of the target.}
\label{fig:Confrontation}
\end{figure}

In summary, let the basic direction $\overrightarrow {OA}$ of the ``rotation mapping'' method be the x-axis. Meanwhile, we use the first two measurement points as the rotation basis. Therefore, the processing result of the sample sparseness confrontation is shown in Fig. \ref{fig:Confrontation}.

In Fig. \ref{fig:Confrontation}, it is shown that the processing result of the sample sparseness is a ``broom'' shape for data samples. The basic motion characteristics of the track are extracted and can applied to the supervised learning's training.

\subsubsection{Extracting the Input Features}\label{C.2}
In Section~\ref{C.1}, in order to address the sample sparseness problem and better extract the input features, the ``sliding window'', the ``rotation mapping'' and the ``relative measurement'' methods are performed on ${\bf{z}}_{1:k}^j$.

\textbf{Step 1} -- \emph{Sliding window (the first time)}: Suppose that the $j$-th track's time is $T$, the measurement sequence of the target is ${\bf{z}}_{1:T}^j = \left\{ {{\bf{z}}_i^j,i = 1, \ldots ,T} \right\}$. The sliding window length is set to $\tau$. Thus, the $j$-th track is processed as follows
\begin{align}\label{eq13}
{\bf{z}}_{1:T}^j \Rightarrow \left\{ \begin{array}{l}
{\bf{z}}_{\tau \left( 1 \right)}^j = \left\{ {{\bf{z}}_i^j,i = 1,2, \ldots ,\tau } \right\}\\
{\bf{z}}_{\tau \left( 2 \right)}^j = \left\{ {{\bf{z}}_i^j,i = 2,3, \ldots ,\tau  + 1} \right\}\\
 \vdots \\
{\bf{z}}_{\tau \left( {T - \tau  + 1} \right)}^j = \left\{ {{\bf{z}}_i^j,i = \left( {T - \tau  + 1} \right), \ldots ,T} \right\}
\end{array} \right..
\end{align}

After the formula \eqref{eq13}, the $j$-th track is intercepted as $( {T - \tau  + 1})$ tracks of the same length $\tau$. It is effective for extracting the basic motion information of the track ``early time'' and ``late time''.

\textbf{Step 2} -- \emph{ Rotation mapping}: Herein, the basic direction $\overrightarrow {OA}$ in Fig. \ref{fig:rotation} is set to the x-axis (i.e., ${\theta _A}{=0}$). Therefore, the rotation angle is $\alpha = - {\theta _B}$. Use the measurements of the first two times as the rotation basis, thus the rotation angle of the $j$-th measurement track is
\begin{align}\label{eq14}
{\alpha ^j} =  - {\tan ^{ - 1}}\left( {\frac{{{\bf{z}}_2^j\left[ y \right] - {\bf{z}}_1^j\left[ y \right]}}{{{\bf{z}}_2^j\left[ x \right] - {\bf{z}}_1^j\left[ x \right]}}} \right),
\end{align}
where ${\bf{z}}_k^j\left[ y \right]$ and ${\bf{z}}_k^j\left[ x \right]$ represent the measurement distance in the y direction and the x direction, respectively.

After the ``sliding window'' method, each track's length is a fixed $\tau$. The vector formed by the measurement of the first two times is rotated to the x-axis direction. Thus, the measurements of the remaining times are then rotated by the same angle ${\alpha ^j}$ using \eqref{eq9}, \eqref{eq10} and \eqref{eq14}, which can be described as
\begin{align}\label{eq15}
{\bf{\mathord{\buildrel{\lower3pt\hbox{$\scriptscriptstyle\smile$}}
\over z} }}_k^j = \left\{ \begin{array}{l}
{\bf{z}}_k^j,k = 1\\
R({\bf{z}}_k^j - {\bf{z}}_{k - 1}^j,\alpha ) + {\bf{\mathord{\buildrel{\lower3pt\hbox{$\scriptscriptstyle\smile$}}
\over z} }}_{k - 1}^j,k = 2,3, \ldots ,\tau
\end{array} \right.,
\end{align}
where ${\bf{\mathord{\buildrel{\lower3pt\hbox{$\scriptscriptstyle\smile$}}
\over z} }}_k^j$ is the new measurement after ``rotation mapping'', $R(\overrightarrow t ,\alpha )$ denotes the vector $\overrightarrow t$ rotated by $\alpha$ ($\alpha>0$ represents counterclockwise and $\alpha<0$ represents clockwise).

Therefore, we can obtain the new measurement ${\bf{\mathord{\buildrel{\lower3pt\hbox{$\scriptscriptstyle\smile$}}
\over z} }}_{1:\tau }^j = \left\{ {{\bf{\mathord{\buildrel{\lower3pt\hbox{$\scriptscriptstyle\smile$}}
\over z} }}_i^j,i = 1, \ldots ,\tau } \right\}$ after the ``rotation mapping'' in \eqref{eq15}. Based on the ``rotation mapping'', the direction influence is removed and all the tracks can be trained in the same direction.

\textbf{Step 3} -- \emph{Sliding window (the second time)}: Then execute the ``sliding window'' method again on ${\bf{\mathord{\buildrel{\lower3pt\hbox{$\scriptscriptstyle\smile$}}\over z} }}_{1:\tau }^j$ as follows
\begin{align}\label{eq16}
 \begin{split}
{\bf{Z}}_{k,\tau }^j &= \underbrace {\left\{ {{{\rm N}_a},{{\rm N}_a}, \ldots } \right\}}_{\tau  - k} + \underbrace {\left\{ {{\bf{\mathord{\buildrel{\lower3pt\hbox{$\scriptscriptstyle\smile$}}
\over z} }}_i^j,i = 1, \ldots ,k - 1,k} \right\}}_k\\
 &= \left\{ {{\bf{\mathord{\buildrel{\lower3pt\hbox{$\scriptscriptstyle\smile$}}
\over z} }}_m^j,m = 1, \ldots ,\tau  - 1,\tau } \right\},
\end{split}
\end{align}
where ${\rm N}_a$ represents the missing value, because the shortages are supplemented by the missing value (some supervised learning algorithms can automatically processing the missing value, such as XGBoost in this paper).

It is indicated in \eqref{eq16} that the $\tau$ nearest measurements are truncated from ${\bf{\mathord{\buildrel{\lower3pt\hbox{$\scriptscriptstyle\smile$}}
\over z} }}_{1:\tau }^j$.  Therefore, a fixed-length feature information is constructed for each sample, which ensures that the ``information'' of the sample are fully expressed.

\textbf{Step 4} -- \emph{Relative measurement}: Use the ``relative measurement'' method to convert ${\bf{Z}}_{k,\tau }^j$ to ${\bf{\tilde Z}}_{k,\tau }^j$ as follows
\begin{align}\label{eq17}
{\bf{\tilde Z}}_{k,\tau }^j &= \left\{ {{\bf{\tilde z}}_1^j, \ldots ,{\bf{\tilde z}}_{\tau  - 1}^j} \right\} \notag\\
&= \left\{ {f\left( {{\bf{\mathord{\buildrel{\lower3pt\hbox{$\scriptscriptstyle\smile$}}
\over z} }}_1^j,{\bf{\mathord{\buildrel{\lower3pt\hbox{$\scriptscriptstyle\smile$}}
\over z} }}_\tau ^j} \right), \ldots ,f\left( {{\bf{\mathord{\buildrel{\lower3pt\hbox{$\scriptscriptstyle\smile$}}
\over z} }}_{\tau  - 1}^j,{\bf{\mathord{\buildrel{\lower3pt\hbox{$\scriptscriptstyle\smile$}}
\over z} }}_\tau ^j} \right)} \right\},
\end{align}
where ${\bf{\tilde z}}_i^j = f\left( {{\bf{\mathord{\buildrel{\lower3pt\hbox{$\scriptscriptstyle\smile$}}
\over z} }}_i^j,{\bf{\mathord{\buildrel{\lower3pt\hbox{$\scriptscriptstyle\smile$}}
\over z} }}_k^j} \right),i \ne k$ denotes the relative distance between the $i$-th new measurement ${\bf{\mathord{\buildrel{\lower3pt\hbox{$\scriptscriptstyle\smile$}}
\over z} }}_i^j$  and the current $k$-th new measurement ${\bf{\mathord{\buildrel{\lower3pt\hbox{$\scriptscriptstyle\smile$}}
\over z} }}_k^j$. Let ${\bf{\mathord{\buildrel{\lower3pt\hbox{$\scriptscriptstyle\smile$}}
\over z} }}_k^j\left[ y \right]$ and ${\bf{\mathord{\buildrel{\lower3pt\hbox{$\scriptscriptstyle\smile$}}
\over z} }}_k^j\left[ x \right]$ represent the new measurement after the ``rotation mapping'' in the y direction and the x direction, respectively. Thus $f\left(  \cdot\right)$ is given by
\begin{align}\label{eq18}
 \begin{split}
f\left( {{\bf{\mathord{\buildrel{\lower3pt\hbox{$\scriptscriptstyle\smile$}}
\over z} }}_i^j,{\bf{\mathord{\buildrel{\lower3pt\hbox{$\scriptscriptstyle\smile$}}
\over z} }}_k^j} \right) &= {\bf{\mathord{\buildrel{\lower3pt\hbox{$\scriptscriptstyle\smile$}}
\over z} }}_i^j - {\bf{\mathord{\buildrel{\lower3pt\hbox{$\scriptscriptstyle\smile$}}
\over z} }}_k^j\\
&= \left( {\begin{array}{*{20}{c}}
{{\bf{\mathord{\buildrel{\lower3pt\hbox{$\scriptscriptstyle\smile$}}
\over z} }}_i^j\left[ x \right] - {\bf{\mathord{\buildrel{\lower3pt\hbox{$\scriptscriptstyle\smile$}}
\over z} }}_k^j\left[ x \right]}\\
{{\bf{\mathord{\buildrel{\lower3pt\hbox{$\scriptscriptstyle\smile$}}
\over z} }}_i^j\left[ y \right] - {\bf{\mathord{\buildrel{\lower3pt\hbox{$\scriptscriptstyle\smile$}}
\over z} }}_k^j\left[ y \right]}
\end{array}} \right),i \ne k.
\end{split}
\end{align}

The relative motion information is extracted after the ``relative measurement''. Therefore, all input feature elements only store relative displacement, which represents the movement change information of each sample. Further, it can be applied to more data ranges.

After the preprocessing of measurements using the above-mentioned ``sliding window'', ``rotation mapping'' and ``relative measurement'' methods, each sample has input features representing the basic motion information of it. Thus, the input features extraction is completed.

\subsubsection{Extracting the Output Variables}\label{C.3}
Similarly, the first step is still to intercept true tracks of different lengths. More specifically, for the $j$-th true track, the method of \eqref{eq13} is still used for processing as follows
\begin{align}\label{eq19}
{\bf{x}}_{1:T}^j \Rightarrow \left\{ \begin{array}{l}
{\bf{x}}_{\tau \left( 1 \right)}^j = \left\{ {{\bf{x}}_i^j,i = 1,2, \ldots ,\tau } \right\}\\
{\bf{x}}_{\tau \left( 2 \right)}^j = \left\{ {{\bf{x}}_i^j,i = 2,3, \ldots ,\tau  + 1} \right\}\\
 \vdots \\
{\bf{x}}_{\tau \left( {T - \tau  + 1} \right)}^j = \left\{ {{\bf{x}}_i^j,i = \left( {T - \tau  + 1} \right), \ldots ,T} \right\}
\end{array} \right..
\end{align}

Since the measurement is executed with the "rotation mapping" when constructing the input feature, it is necessary to perform the "rotation mapping" on the true state before training. However, if the ``rotation mapping'' is performed directly on the true track, the relative error between the true state and the measurement will be changed. Meanwhile, the error of the training sample will increase. Therefore, this paper adopts another way to operate. More specifically, The error vector ${\bf{r}}_k^j$ is rotated by the angle ${\alpha ^j}$ corresponding to the $j$-th track as follows
\begin{align}\label{eq20}
{\bf{\mathord{\buildrel{\lower3pt\hbox{$\scriptscriptstyle\smile$}}
\over r} }}_k^j = R\left( {{\bf{r}}_k^j,{\alpha ^j}} \right) = R\left( {{\bf{x}}_k^j - {\bf{z}}_k^j,{\alpha ^j}} \right),k = 1,2, \ldots ,\tau,
\end{align}
where ${\bf{\mathord{\buildrel{\lower3pt\hbox{$\scriptscriptstyle\smile$}}
\over r} }}_k^j$ is the error after the ``rotation mapping''.

It can be known from \eqref{eq9} and \eqref{eq10} that the rotation operation will not change the vector's length. Thus, the error between the true state and the original measurement will not change, ensuring the information consistency of the sample data. Further, using ${\bf{\mathord{\buildrel{\lower3pt\hbox{$\scriptscriptstyle\smile$}}
\over r} }}_k^j$ as the output variable of each sample, the output variables are constructed.

\subsection{SLF Training Phase}\label{Sec: analysis-D}
Suppose we have collected the true state ${\bf{x}}_k^j$ and the historical measurement sequence ${\bf{z}}_{1:k}^j$. We have $N$ tracks, each track from $k = 1$ to $k = T$. Therefore, the data set is represented as
\begin{align}\label{eq21}
\Phi  = \left\{ {\left( {{\bf{z}}_{1:k}^i,{\bf{x}}_k^i} \right),i = 1,...,N,k = 1,...,T} \right\}.
\end{align}

For $\Phi $, we first perform Section~\ref{Sec: analysis-C} to effectively extract the basic motion information. Then we complete the output variable construction by \eqref{eq21}.

After the preprocessing of data set $\Phi $, the input features ${\bf{\tilde Z}}_{k,\tau }^j$ and output variables ${\bf{\mathord{\buildrel{\lower3pt\hbox{$\scriptscriptstyle\smile$}}
\over r} }}_k^j$ can be obtained. Thus, based on Section~\ref{Sec: analysis-A}, we can construct the representation of ``hidden model'' -- the hypothesis function $h$, which can be described as
\begin{align}\label{eq22}
h\left( {{\bf{\tilde Z}}_{k,\tau }^j,\phi } \right):{\bf{\tilde Z}}_{k,\tau }^j \to {\bf{\mathord{\buildrel{\lower3pt\hbox{$\scriptscriptstyle\smile$}}
\over r} }}_k^j,
\end{align}
where $\phi$ is the parameter set of $h$. The above \eqref{eq22} represents the hypothesis function $h$ establishes a mapping relationship between ${\bf{\tilde Z}}_{k,\tau }^j$ and ${\bf{\mathord{\buildrel{\lower3pt\hbox{$\scriptscriptstyle\smile$}}
\over r} }}_k^j$.

We hope that the state estimation ${\bf\hat{\mathord{\buildrel{\lower3pt\hbox{$\scriptscriptstyle\smile$}}
\over r} }}_k^j$ output by the function $h$ can approximate the output variables ${\bf{\mathord{\buildrel{\lower3pt\hbox{$\scriptscriptstyle\smile$}}\over r} }}_k^j$ as much as possible. Therefore, the loss function $J\left( \phi  \right)$ is used to measure the quality of the hypothesis function's parameter $\phi$. Then the minimal loss function is used to find the optimal parameter $\hat \phi$ as follows
\begin{align}\label{eq23}
\hat \phi  = \mathop {\arg \min }\limits_\phi  \sum\limits_{j = 1}^S {\sum\limits_{k = 1}^\tau  {l\left( {{\bf{\mathord{\buildrel{\lower3pt\hbox{$\scriptscriptstyle\smile$}}
\over r} }}_k^j,h\left( {{\bf{\tilde Z}}_{k,\tau }^j,\phi } \right)} \right)} }  + \lambda \Omega \left( \phi  \right),
\end{align}
where $S = N\left( {T - \tau  + 1} \right)$ denotes the total number of tracks after data preprocessing.

There are many options for the training error term $l\left(  \cdot  \right)$ in the loss function. For the filtering problem of target tracking, since it is a regression problem, we choose the training error term as the root mean square error (RMSE), which is given by
\begin{align}\label{eq24}
l\left( {{\bf{\mathord{\buildrel{\lower3pt\hbox{$\scriptscriptstyle\smile$}}
\over r} }}_k^j,h\left( {{\bf{\tilde Z}}_{k,\tau }^j,\phi } \right)} \right) = \sqrt {\frac{1}{G}\sum\limits_{j = 1}^S {\sum\limits_{k = 1}^\tau  {\left\| {{\bf{\mathord{\buildrel{\lower3pt\hbox{$\scriptscriptstyle\smile$}}
\over r} }}_k^j - h\left( {{\bf{\tilde Z}}_{k,\tau }^j,\phi } \right)} \right\|_{\rm{2}}^{\rm{2}}} } } ,
\end{align}
where $G = N\left( {T - \tau  + 1} \right) * \tau$ is the total number of samples (points).

Based on \eqref{eq23} and \eqref{eq24}, iterate continuously and train to obtain a machine model network. Therefore, use the optimal parameter $\hat \phi$ to get the optimal estimation ${\bf\hat{\mathord{\buildrel{\lower3pt\hbox{$\scriptscriptstyle\smile$}}
\over r} }}_k^j = h\left( {{\bf{\tilde Z}}_{k,\tau }^j,\hat \phi } \right)$. But ${\bf\hat{\mathord{\buildrel{\lower3pt\hbox{$\scriptscriptstyle\smile$}}
\over r} }}_k^j$ is the error obtained after the ``rotation mapping''. Thus, it needs to be processed by the ``inverse rotation mapping'' and the ``inverse transformation'' below to get the state estimation ${\bf{\hat x}}_k^j$.

\subsection{SLF Estimation and Application Phase}\label{Sec: analysis-E}
Based on the above-mentioned steps, we can obtain the hypothesis function $h$ and its optimal parameter $\hat \phi$ in the minimal loss function with the training data. Thus, the ``hidden model'' is successfully constructed. Next, the state estimation is started. More specifically, the new measurement is mapped to obtain a new filtering value.

First, create the input features ${\bf{\tilde Z}}_{k,\tau }^{j,{\rm{n}}}$ for the new measurement ${\bf{z}}_{1:k}^{j,{\rm{n}}}$ using \eqref{eq14} -- \eqref{eq17}, where ``$\rm{n}$'' stands for new. Then, put ${\bf{\tilde Z}}_{k,\tau }^{j,{\rm{n}}}$ into the trained hypothesis function $h$ for application as follows
\begin{align}\label{eq25}
{\bf\hat{\mathord{\buildrel{\lower3pt\hbox{$\scriptscriptstyle\smile$}}
\over r} }}_k^{j,{\rm{n}}} = h\left( {{\bf{\tilde Z}}_{k,\tau }^{j,{\rm{n}}},\hat \phi } \right).
\end{align}

Next, perform the ``inverse rotation mapping'' on the new error ${\bf\hat {\mathord{\buildrel{\lower3pt\hbox{$\scriptscriptstyle\smile$}}
\over r} }}_k^{j,{\rm{n}}}$ using \eqref{eq11} and \eqref{eq12}. Thus, the filtering error ${\bf{\hat r}}_k^{j,{\rm{n}}}$ in the original coordinate dimension is given by
\begin{align}\label{eq26}
{\bf{\hat r}}_k^{j,{\rm{n}}} = {R^{ - {\rm{1}}}}\left( {{\bf\hat {\mathord{\buildrel{\lower3pt\hbox{$\scriptscriptstyle\smile$}}
\over r} }}_k^{j,{\rm{n}}},{\alpha ^j}} \right),
\end{align}
where ${R^{ - {\rm{1}}}}\left(  \cdot  \right)$ is the ``inverse rotation mapping'' obtained by \eqref{eq11} and \eqref{eq12}.

Finally, the ``inverse transformation'' is performed on ${\bf{\hat r}}_k^{j,{\rm{n}}}$ in the original coordinate dimension to get the state estimation. Herein, the ``inverse transformation'' can be described as
\begin{align}\label{eq27}
{\bf{\hat x}}_k^{j,{\rm{n}}} = {\bf{\hat r}}_k^{j,{\rm{n}}} + {\bf{z}}_k^{j,{\rm{n}}}.
\end{align}

Based on this phase, the ``hidden model'' is mapped to form corresponding state estimation. Thus, the SLF can filter the new measurement and obtain the new state estimation ${\bf{\hat x}}_k^{j,{\rm{n}}}$.

\begin{figure*}[htbp]
\centering
\includegraphics[width=1\textwidth]{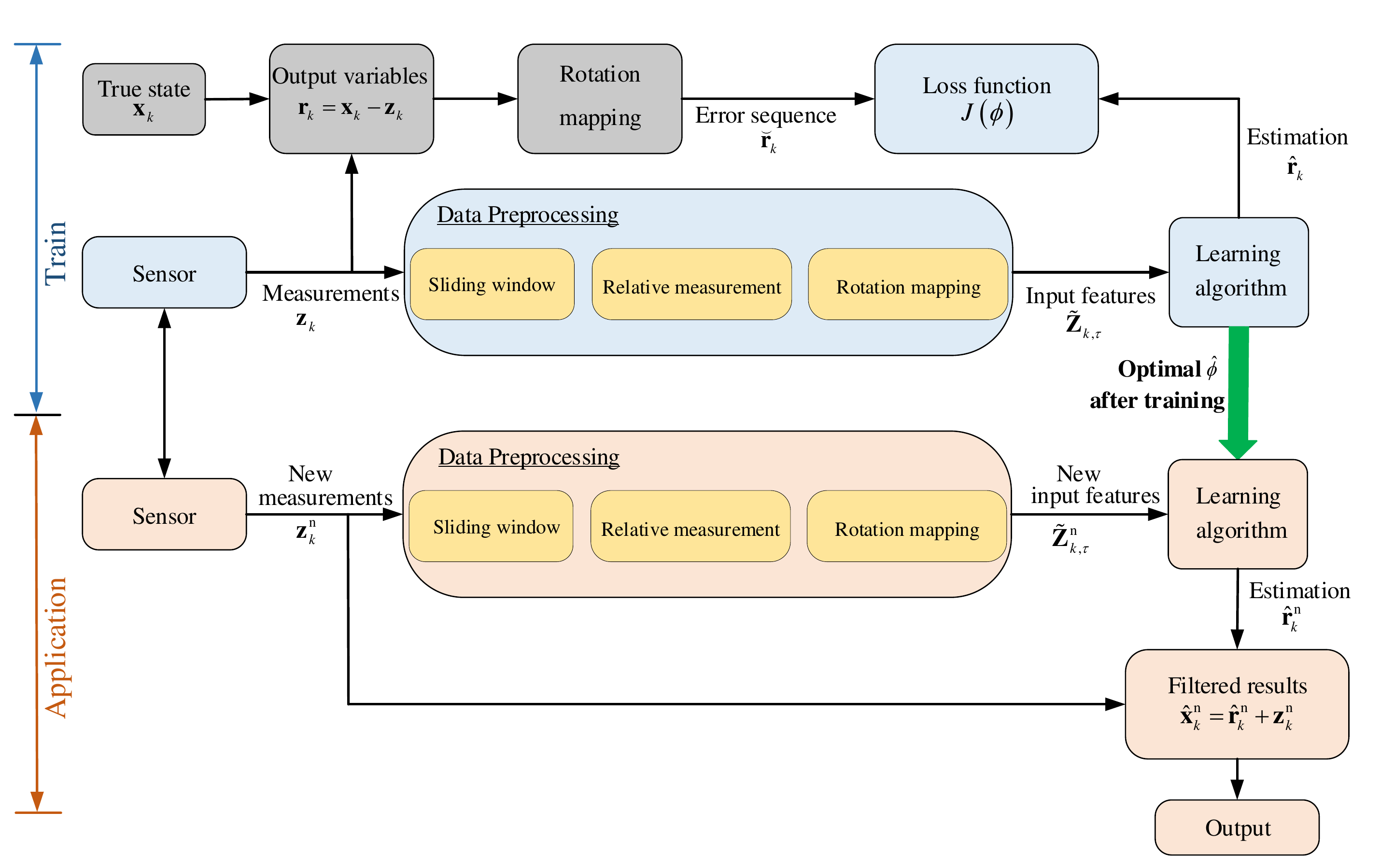}\\
\caption{The SLF framework is mainly divided into two parts, one is to use the measurement and the true state to train the ``hidden model'', the second is to predict the new measurement using the trained model.}\label{fig:SLF}
\end{figure*}

According to the above-mentioned phases, we finally give an overall framework for SLF as shown in Fig. \ref{fig:SLF}.

\section{XGBoost Implementation} \label{Sec: XGBoost}

In 1999, Friedman derived a gradient descent based boosting method, which is called gradient tree boosting (GTB) \cite{Friedman2001}. Meanwhile, GTB uses classification and regression trees (CART) \cite{Bhargava2017} as the base classifier. XGBoost is proposed by Chen Tianqi in recent years, which is a concrete implementation of GTB and is an efficient and powerful open source boosted tree toolkit \cite{Tianqi2016}. XGBoost has excellent performance in many fields, such as classification \cite{Zhang2018}, prediction \cite{Zhong2018} and regression \cite{Gumus2017}.

Herein, the corresponding Section~\ref{Sec: analysis-C} is performed on the training data. Then, XGBoost is selected to implement SLF as a training algorithm.

\textbf{Step 1} -- \emph{The XGBoost's hypothetical function}: Because XGBoost is a tree-based algorithm, the parameter $\phi$ of $h$ consists of two parts: one is the tree's structure, the other is the score of each leaf node \cite{Tianqi2016}. Therefore, it is need to find the optimal tree's structure and the corresponding optimal leaf node by minimizing the loss function.

\textbf{Step 2} -- \emph{The XGBoost's loss function}: The XGBoost's specific theory can be found in \cite{Tianqi2016}. Herein, we only focus on the XGBoost training for the hypothesis function $h$ under the SLF framework. XGBoost is based on a set of CARTs for learning. Therefore, for the filtering problem of target tracking, a set of tree models are represented as follows
\begin{align}\label{eq28}
{\bf\hat{\mathord{\buildrel{\lower3pt\hbox{$\scriptscriptstyle\smile$}}
\over r} }}_k^j = \sum\limits_{m = 1}^M {{f_m}\left( {{\bf{\tilde Z}}_{k,\tau }^j} \right)} ,{f_m} \in \Gamma,
\end{align}
where $M$ is the number of trees, $\Gamma$ denotes all possible CARTs, ${f_m}$ is a specific CART, ${\bf{\tilde Z}}_{k,\tau }^j$ denotes the input features under the filtering problem of target tracking.

For the filtering problem of $N$ tracks with time $T$, the total number of samples (points) is $G$ in \eqref{eq24}. To better explain the mathematical theory, we rewrite ${\bf{\tilde Z}}_{k,\tau }^j$ as ${{\bf{\tilde Z}}_{g,\tau }}$, ${\bf\hat{\mathord{\buildrel{\lower3pt\hbox{$\scriptscriptstyle\smile$}}
\over r} }}_k^j$ as ${\bf\hat{\mathord{\buildrel{\lower3pt\hbox{$\scriptscriptstyle\smile$}}
\over r} }}_g$, and ${\bf{\mathord{\buildrel{\lower3pt\hbox{$\scriptscriptstyle\smile$}}
\over r} }}_k^j$ as ${{\bf{\mathord{\buildrel{\lower3pt\hbox{$\scriptscriptstyle\smile$}}
\over r} }}_g}$, where $g = 1,2,...G$.

Therefore, the general loss function in XGBoost is
\begin{align}\label{eq29}
J = \sum\limits_{g = 1}^G {l\left( {{{{\bf{\mathord{\buildrel{\lower3pt\hbox{$\scriptscriptstyle\smile$}}
\over r} }}}_g},{{{\bf\hat{\mathord{\buildrel{\lower3pt\hbox{$\scriptscriptstyle\smile$}}
\over r} }}_g}}} \right) + \sum\limits_{m = 1}^M {\Omega \left( {{f_m}} \right)} }.
\end{align}

Since the tree model is an addition model, we use a greedy strategy with a forward distribution algorithm. At step $t$, we add an optimal CART ${f_t}$ (i.e., the tree that minimizes the loss function on the basis of the existing $t-1$ trees). Therefore, ${f_t}$ is given by
\begin{align}\label{eq30}
{J^{(t)}} &= \sum\limits_{g = 1}^G {l\left( {{{{\bf{\mathord{\buildrel{\lower3pt\hbox{$\scriptscriptstyle\smile$}}
\over r} }}}_g},{\bf\hat{\mathord{\buildrel{\lower3pt\hbox{$\scriptscriptstyle\smile$}}
\over r} }}_g^{(t)}} \right)}  + \sum\limits_{i = 1}^t {\Omega \left( {{f_i}} \right)} \notag \\
&= \sum\limits_{g = 1}^G {l\left( {{{{\bf{\mathord{\buildrel{\lower3pt\hbox{$\scriptscriptstyle\smile$}}
\over r} }}}_g},{\bf\hat{\mathord{\buildrel{\lower3pt\hbox{$\scriptscriptstyle\smile$}}
\over r} }}_g^{(t-1)} + {f_t}\left( {{{{\bf{\tilde Z}}}_{g,\tau }}} \right)} \right)}  + \Omega \left( {{f_t}} \right) + {\rm{C}},
\end{align}
where ${\bf\hat{\mathord{\buildrel{\lower3pt\hbox{$\scriptscriptstyle\smile$}}
\over r} }}_g^{(t)}$ is the output of the $t$-th tree (the error between the true state and the measurement after ``rotation mapping''), ${\rm{C}}$ denotes the complexity of the previous $t-1$ trees.

Next, we perform the second-order Taylor expansion on \eqref{eq30}. Meanwhile, because our goal is to minimize the loss function ${J^{(t)}}$ with the variable ${f_t}\left( {{{{\bf{\tilde Z}}}_{g,\tau }}} \right)$, the constant term ${\rm{C}}$ can be removed. Thus, we can get
\begin{align}\label{eq31}
{J^{(t)}} \approx \sum\limits_{g = 1}^G {\left[ {{e_g}{f_t}\left( {{{{\bf{\tilde Z}}}_{g,\tau }}} \right) + \frac{1}{2}{h_g}f_t^2\left( {{{{\bf{\tilde Z}}}_{g,\tau }}} \right)} \right]}  + \Omega \left( {{f_t}} \right),
\end{align}
where
\begin{align}\label{eq32}
\left\{ \begin{array}{l}
{e_g} = {\partial _{{\bf\hat{\mathord{\buildrel{\lower3pt\hbox{$\scriptscriptstyle\smile$}}
\over r} }}_g^{(t - 1)}}}l\left( {{{{\bf{\mathord{\buildrel{\lower3pt\hbox{$\scriptscriptstyle\smile$}}
\over r} }}}_g},{\bf\hat{\mathord{\buildrel{\lower3pt\hbox{$\scriptscriptstyle\smile$}}
\over r} }}_g^{(t - 1)}} \right)\\
{h_g} = \partial _{{\bf\hat{\mathord{\buildrel{\lower3pt\hbox{$\scriptscriptstyle\smile$}}
\over r} }}_g^{(t - 1)}}^2l\left( {{{{\bf{\mathord{\buildrel{\lower3pt\hbox{$\scriptscriptstyle\smile$}}
\over r} }}}_g},{{\bf\hat{\mathord{\buildrel{\lower3pt\hbox{$\scriptscriptstyle\smile$}}
\over r} }}_g^{(t - 1)}}} \right)
\end{array} \right..
\end{align}

For the training error term $l\left(  \cdot  \right)$, we also select the root mean square error (RMSE) according to \eqref{eq24}.

For the regularization term $\Omega \left( {{f_t}} \right)$, XGBoost chooses it in \cite{Tianqi2016} is
\begin{align}\label{eq33}
\Omega \left( {{f_t}} \right) = \gamma P + \frac{1}{2}\lambda \sum\limits_{j = 1}^P {w_j^2},
\end{align}
where $P$ is the number of leaf nodes of the tree, the values of the $P$ leaf nodes form a $P$ dimensional vector $w$, $\gamma$ and $\lambda$ represent parameters and are manually set. Obviously, the larger $\gamma$ is, the more we hope to obtain a simple tree.

More specifically, for how to obtain the optimal tree structure and the corresponding optimal leaf nodes, please refer to \cite{Tianqi2016}. Due to paper space limitations, this paper will not make specific derivations. In the end, XGBoost obtained the optimal tree structure and corresponding optimal leaf nodes by training. In other words, we can find the optimal parameter $\hat \phi$ by minimizing the loss function.

\section{Simulation and Results} \label{Sec: simulation}

\subsection{Training Data Generation}\label{Sec: simulation-A}

Based on Section~\ref{Sec: XGBoost} and the SLF framework, we further proposed a specific implementation method: a XGBoost based online tracking filter (XGBF).

Because the existing data set suitable for XGBF and KF cannot be found, this paper uses specific models to generate the data set for simulation experiments. At the same time, this paper mainly considers the single filtering problem of target tracking. Herein, the measurement correlation problem in multi-target environments and the clutter problem are not considered.

For the linear system, we select a scene where the track is transformed according to the constant velocity (CV) model \cite{Yaakov2001}. Next, we use this scene to generate a data set of XGBF and KF simulation. The state equation is given by
\begin{align}\label{eq34}
{{\bf{x}}_{k + 1}} = {\bf{F}}{{\bf{x}}_k} + {{\bf{w}}_k},
\end{align}
and the state transition function is
\begin{align}\label{eq35}
\mathbf{F}=\mathbf{I}_2\otimes\left[
  \begin{matrix}
  1&\Delta t\\
  0&1\\
  \end{matrix}
  \right],
\end{align}
where ${{\bf{I}}_2}$ denotes the $2 \times 2$ identity matrix, $ \otimes $ is the Kronecker product, the state vector of the target defined as
$\mathbf{x}_k=\begin{bmatrix}x_k & \dot{x}_k & y_k &\dot{y}_k\end{bmatrix}^{\top}$, ${\left[ {{x_k},{y_k}} \right]^{\top}}$ and ${\left[ {{{\dot x}_k},{{\dot y}_k}} \right]^{\top}}$ are the target position and velocity in Cartesian coordinate system, $\left\{ {{{\bf{w}}_k},k \in \mathbb{N}} \right\}$ is an i.i.d process noise sequence, $\Delta t$ is the sensor scanning interval.

The measurement equation is as follows
\begin{align}\label{eq36}
{{\bf{z}}_k} = {\bf{H}}{{\bf{x}}_k} + {{\bf{v}}_k},
\end{align}
and the measurement function is
\begin{align}\label{eq37}
{\bf{H}} = \left[ {\begin{array}{*{20}{c}}
1&0&0&0\\
0&0&1&0
\end{array}} \right],
\end{align}
where $\left\{ {{{\bf{v}}_k},k \in \mathbb{N}} \right\}$ is an i.i.d measurement noise sequence.

Suppose that the process noise and the measurement noise follow the zero-mean Gaussian distribution \cite{Yaakov2001}. The covariances are ${\bf{Q}}$ and ${\bf{R}}$, respectively. Therefore, the covariances can be written as
\begin{align}
\label{eq38}{\bf{Q}} &= {q_s}{{\bf{I}}_2} \otimes \left[ {\begin{array}{*{20}{c}}
{{{\Delta {t^3}} \mathord{\left/
 {\vphantom {{\Delta {t^3}} 3}} \right.
 \kern-\nulldelimiterspace} 3}}&{{{\Delta {t^2}} \mathord{\left/
 {\vphantom {{\Delta {t^2}} 2}} \right.
 \kern-\nulldelimiterspace} 2}}\\
{{{\Delta {t^2}} \mathord{\left/
 {\vphantom {{\Delta {t^2}} 2}} \right.
 \kern-\nulldelimiterspace} 2}}&{\Delta t}
\end{array}} \right],\\
\label{eq39}{\bf{R}} &= \left[ {\begin{array}{*{20}{c}}
{v_x^2}&0\\
0&{v_y^2}
\end{array}} \right],
\end{align}
where ${q_s}$ is the process noise intensity, ${v_x}$ and ${v_y}$ are the standard deviations of the measurement noise in the x-axis direction and the y-axis direction, respectively.

According to the above-mentioned formula, corresponding measurement sequences and true track sequences are generated. They are used as the training set of XGBF after preprocessing, so as to train the ``hidden model''. Subsequently, the state estimation is performed on the new data and compared with KF.
\subsection{The Effects of XGBF Hyper-parameters}

Hyper-parameters have a certain impact on the estimation accuracy of XGBF. For example, if the sliding window length $\tau$ is too small, the history information will be lost too much. However, if $\tau$ is too big, the input features will introduce more noise. Thus, the choice of $\tau$ should be moderate. For another example, the more training samples, the more accurate the data feature extraction will be.

\begin{figure*}[htbp]
    \centering
	  \subfigure[]{
       \includegraphics[width=1.7 in]{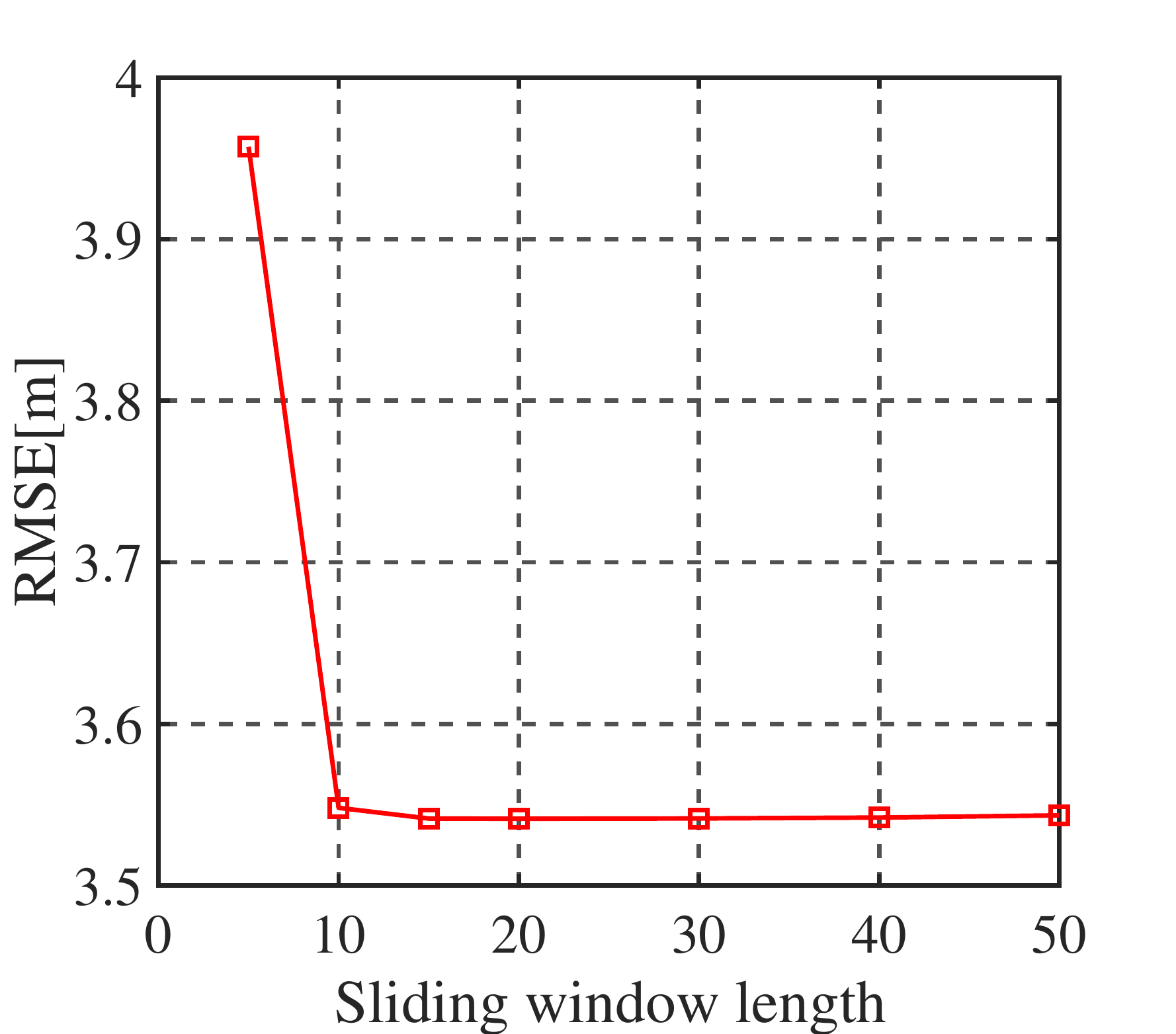}}
    \label{1a}\hfill
	  \subfigure[]{
        \includegraphics[width=1.7 in]{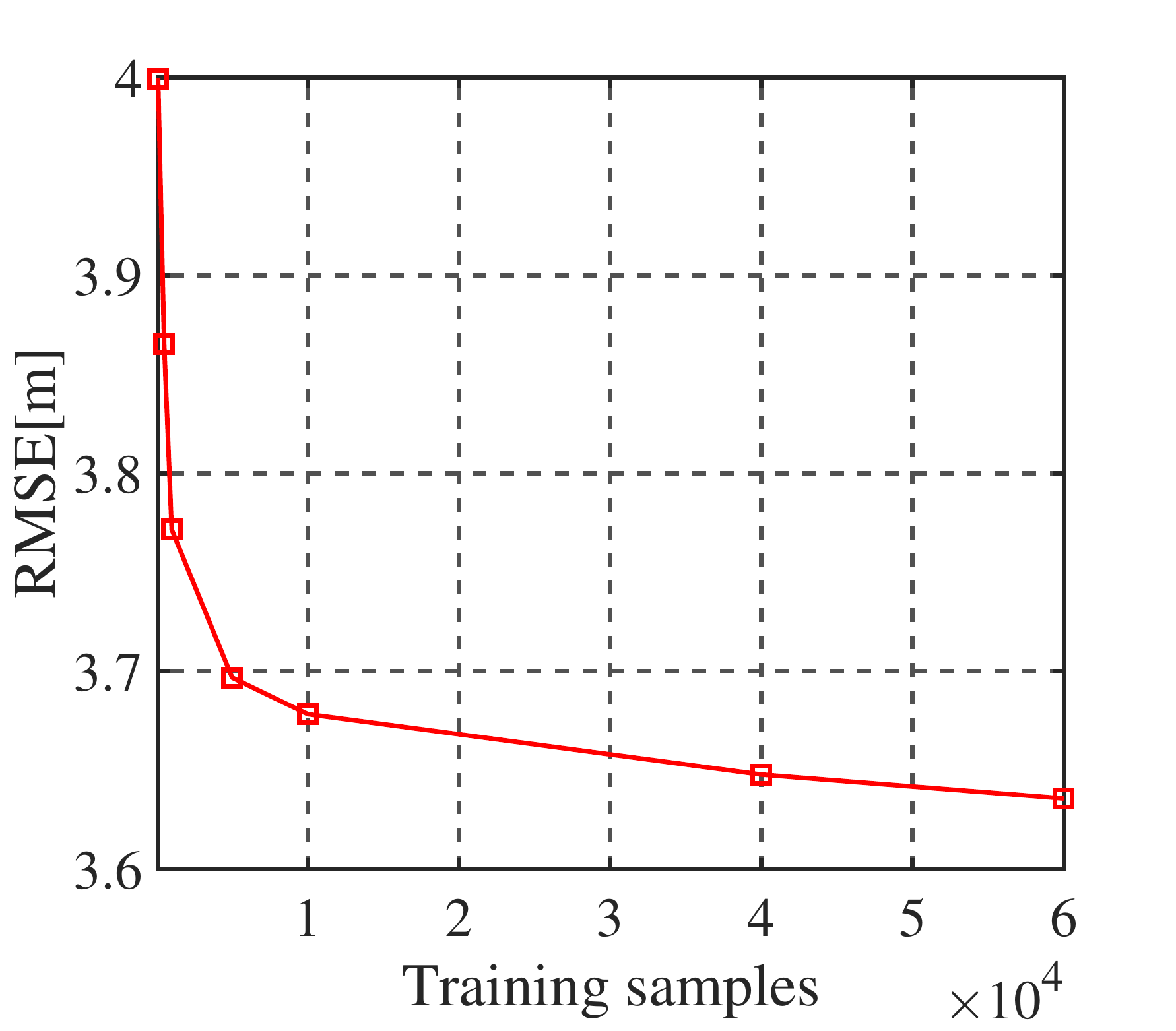}}
    \label{1b}\hfill
	  \subfigure[]{
        \includegraphics[width=1.7 in]{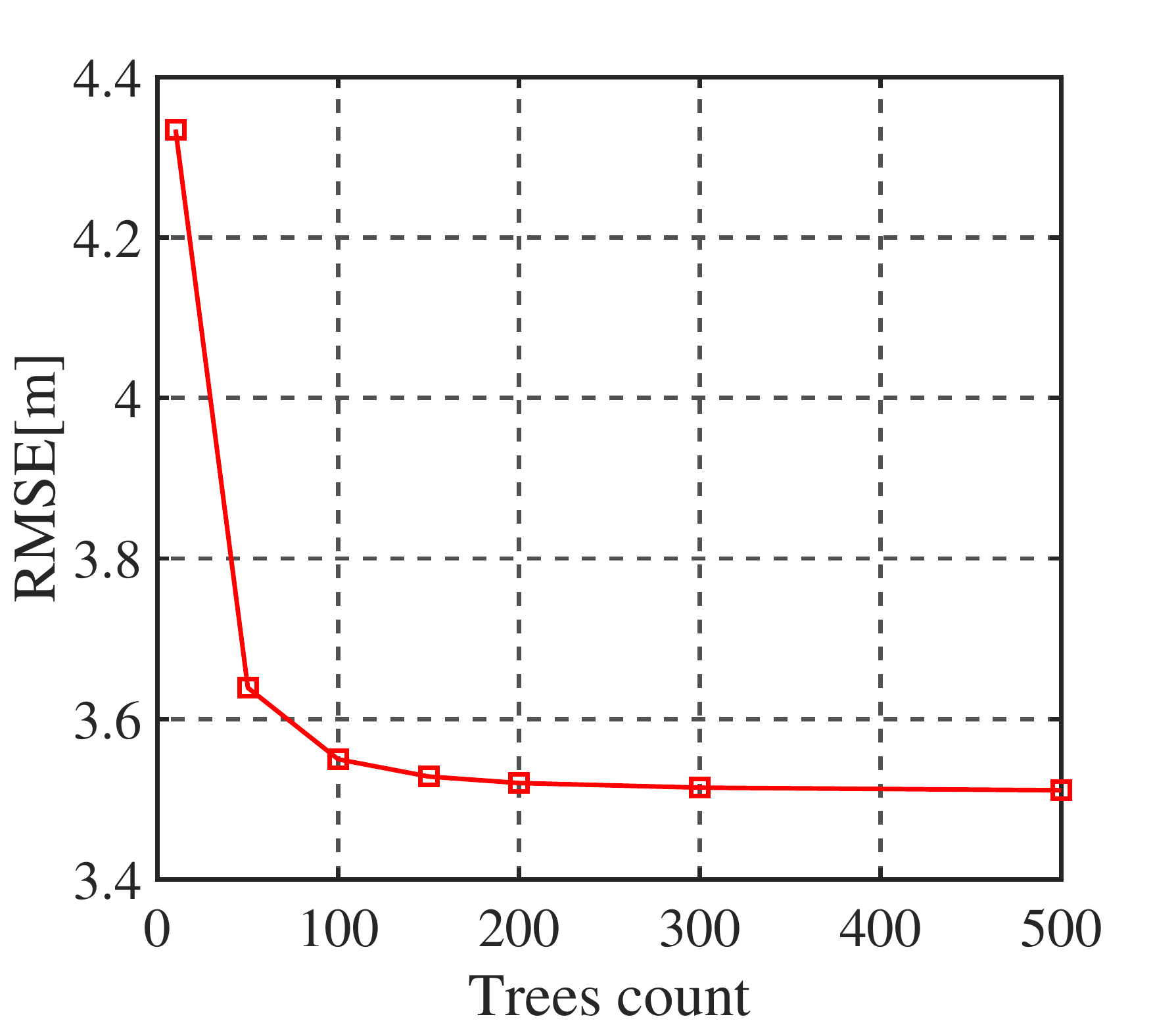}}
    \label{1c}\hfill
	  \subfigure[]{
        \includegraphics[width=1.7 in]{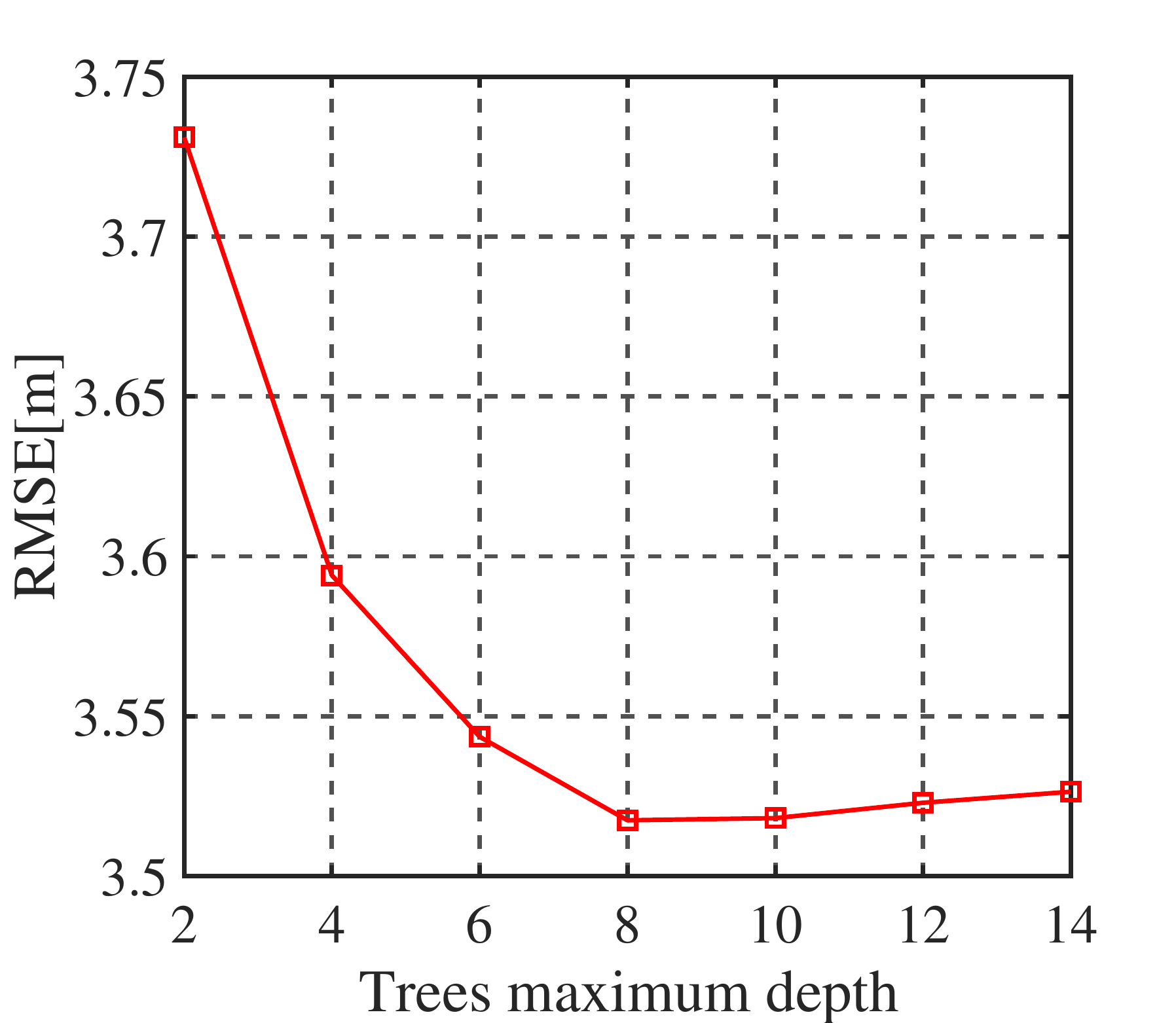}}
     \label{1d}\hfill
	  \caption{(a) The effect of the sliding window length, if the sliding window length is too big, the input features will introduce more noise. (b) The effect of the number of the training samples, the accuracy does not improve much after the training samples reach a certain level. (c) The effect of the trees number. (d) The effect of the trees maximum depth, the larger the value, the more complicated and accurate the model is, but it is easy to over-fitting.}
	  \label{fig:hyperparameters}
\end{figure*}

Therefore, we study the number of XGBoost trees (representing the number of iterations), the sliding window length $\tau$, the number of training samples and the trees maximum depth as follows in Fig. \ref{fig:hyperparameters}. Other hyper-parameters use the default values \cite{Tianqi2016}.

Considering the calculation power and estimation accuracy based on Fig. \ref{fig:hyperparameters}, we set the values of hyper-parameters as shown in Table \ref{table: hyper-parameter}. The learning rate is not plotted, but the characteristic is similar to the trees maximum depth.

\begin{table}[htbp]
		\centering
        \caption{The XGBF hyper-parameters setting}
\begin{tabular}{lcl}
\toprule
\textbf{Parameter}&\textbf{Definition}&\textbf{Value} \\
\midrule
samples& Training samples & 10000 \\
$\tau $& Sliding window length & 20 \\
nrounds& The number of trees & 500 \\
max depth& Maximum depth of a tree & 8 \\
eta& The learning rate & 0.05 \\
\bottomrule
\end{tabular}
\label{table: hyper-parameter}
\end{table}
\subsection{The Effects of ``Sample Sparseness Confrontation'' Method}
\subsubsection{The track after the ``sample sparseness confrontation'' process}
\begin{figure}[htbp]
\centering
\subfigure[]{\includegraphics[width=0.8\columnwidth,draft=false]{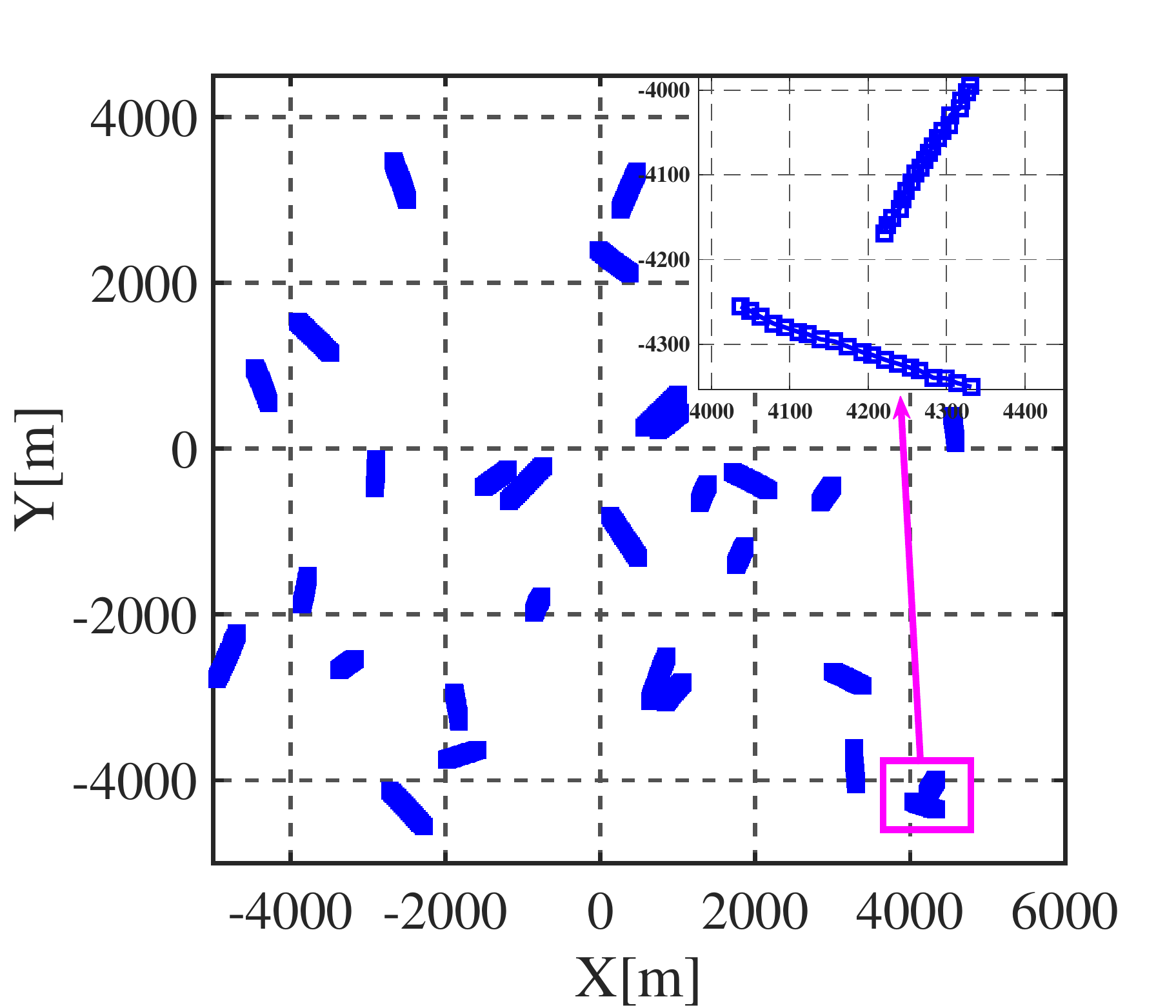}}
\subfigure[]{\includegraphics[width=0.8\columnwidth,draft=false]{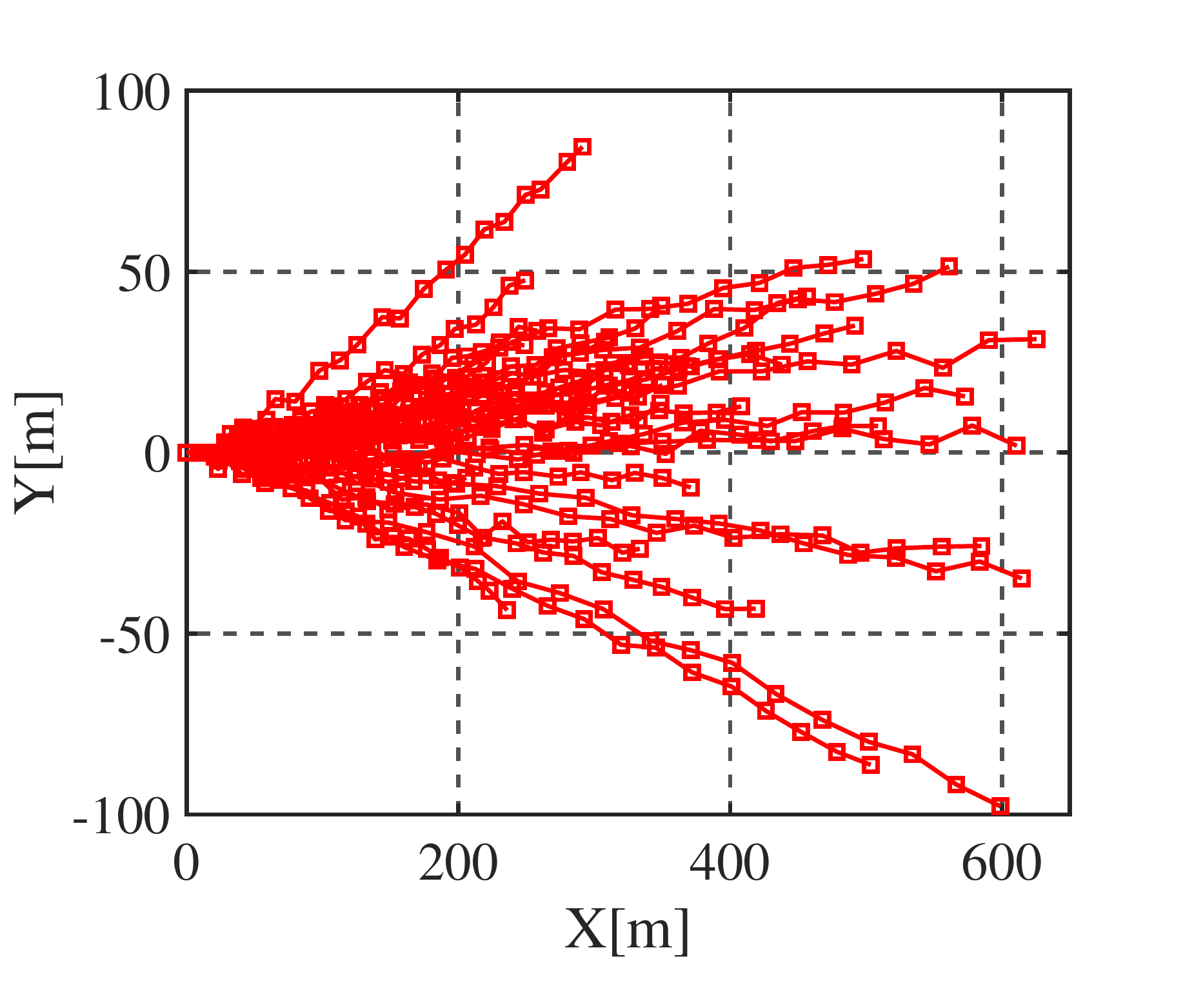}}
\caption{(a) Before the ``sample sparseness confrontation'' processing, the initial state of target is uniformly generated in [$\pm$ 5000m, $\pm$ 25m/s, $\pm$5000m, $\pm$ 30m/s]. (b) After the ``sample sparseness confrontation'' processing, the processing result is a ``broom'' shape, which is beneficial to the algorithm for model training.}
\label{fig:Sample Sparseness}
\end{figure}

Combined with Section~\ref{Sec: analysis-C}, we can get the track after the ``sample sparseness confrontation'' process (only 30 tracks are shown in Fig. \ref{fig:Sample Sparseness})

It can be seen from Fig. \ref{fig:Sample Sparseness} that the simulation results are in accordance with the theoretical effect of Fig. \ref{fig:Confrontation} after the ``sample sparseness confrontation''. Thereby solving the sample sparseness problem(time, space, and angle).

\subsubsection{The actual effect of ``rotation mapping''}
In order to verify the actual effect of ``rotation mapping'' in the ``sample sparseness confrontation'' method, we consider an extreme sample distribution. More specifically, training samples and test samples are in different initial states and directions, but the essential motion characteristics are similar. Therefore, we make the following settings:
\begin{itemize}
  \item The target initial state of training samples is randomly and uniformly generated in [0 ~ 5000m, + 25m/s, 0 ~ 5000m, + 30m/s]. The number of training samples is 10000.
  \item The target initial state of test samples is randomly and uniformly generated in [-5000 ~ 0m, -25m/s, -5000 ~ 0m, -30m/s]. The number of test samples is 5000.
  \item Suppose $\Delta t$ = 1s and $T$ = 30s. The process noise satisfies the Gaussian distribution ${w_k} \sim {{\cal N}}\left( {0,{\bf{Q}}} \right)$, where ${q_s}$ = 1. The measurement noise ${v_k} \sim {{\cal N}}\left( {0,{\bf{R}}} \right)$, where $v_x^2$ = 30 and $v_y^2$ = 20.
\end{itemize}

\begin{figure}[htbp]
\centering
{\includegraphics[width=0.8\columnwidth,draft=false]{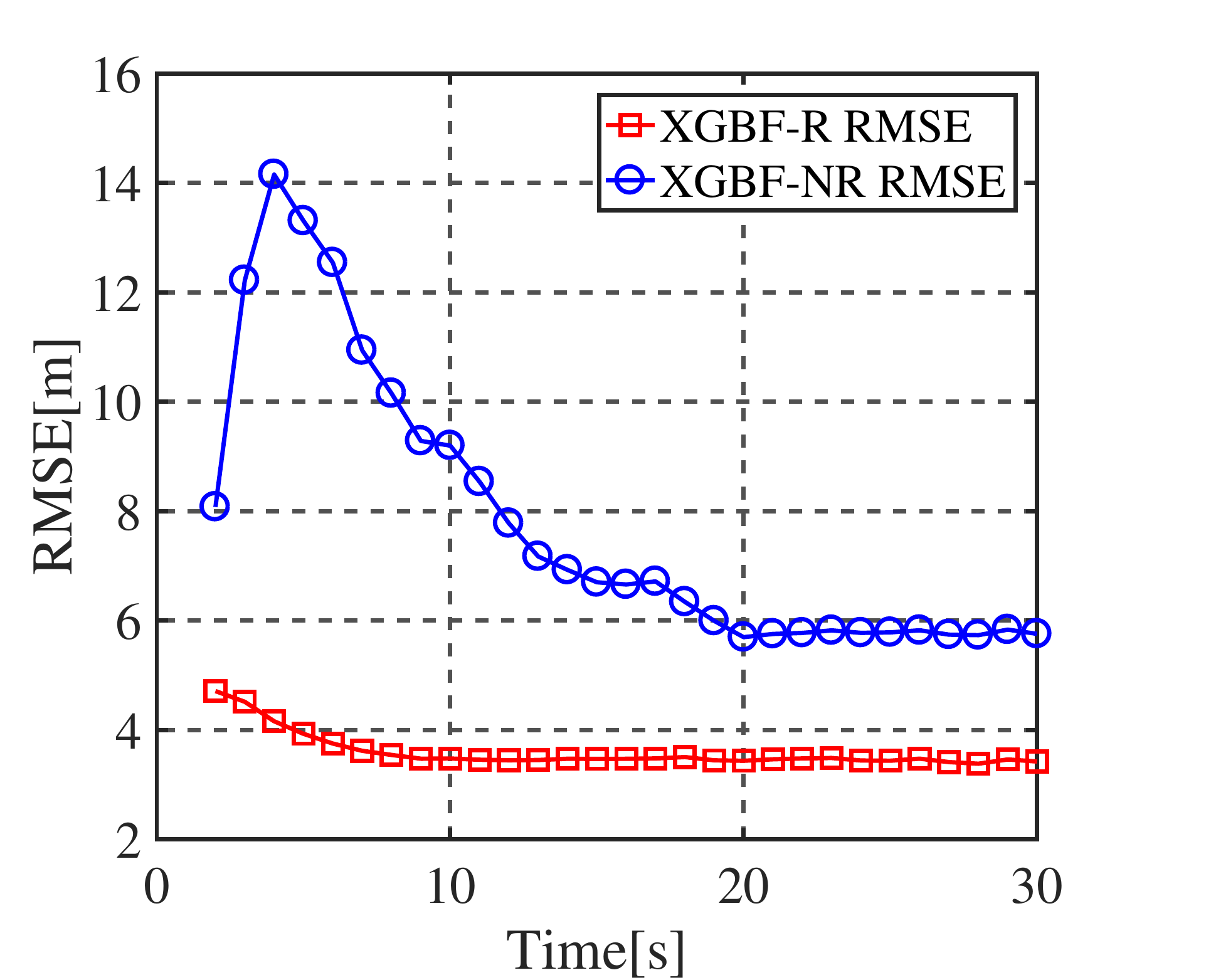}}
\caption{The estimation accuracy of XGBF with the ``rotation mapping'' (XGBF-R) and XGBF without the ``rotation mapping'' (XGBF-NR)}
\label{fig:duibi}
\end{figure}

Based on the above-mentioned settings, we use XGBF with the ``rotation mapping'' and XGBF without the ``rotation mapping'' to make predictions. Thus, the simulation results are shown in Fig. \ref{fig:duibi}.

It is indicated in Fig. \ref{fig:duibi} that the ``sample sparseness confrontation'' method has very good results in extreme cases. After processing using the ``sample sparseness confrontation'', whether it is a normal distributed sample or an extreme distributed sample, XGBF can always achieve a good estimation accuracy. This is also the main reason for us to adopt the ``sample sparseness confrontation''.
\subsection{Performance Comparison}

\subsubsection{KF VS. XGBF in the ``General Case''}

In the ``general case'', use Section~\ref{Sec: simulation-A} to perform the training and test set generation. Suppose $\Delta t$ = 1s and $T$ = 50s. The process noise satisfies the Gaussian distribution ${w_k} \sim {{\cal N}}\left( {0,{\bf{Q}}} \right)$, where ${q_s}$ = 1. The measurement noise ${v_k} \sim {{\cal N}}\left( {0,{\bf{R}}} \right)$, where $v_x^2$ = 30 and $v_y^2$ = 20. For XGBF, choose 10000 training samples and 5000 test samples. For KF, the tracks number $N$ = 5000 (i.e., 5000 Monte Carlo). Meanwhile, the parameters of XGBF are set from Table \ref{table: hyper-parameter}. In addition, RMSE is selected as the estimation accuracy measure of XGBF and KF.

In the actual filtering problem of target tracking, the initial state is generally random. Therefore, we consider the target has the random initial state. Herein, the initial state is uniformly generated in [$\pm$ 5000m, $\pm$ 25m/s, $\pm$5000m, $\pm$ 30m/s].
\begin{figure}[htbp]
\centering
\subfigure[]{\includegraphics[width=0.8\columnwidth,draft=false]{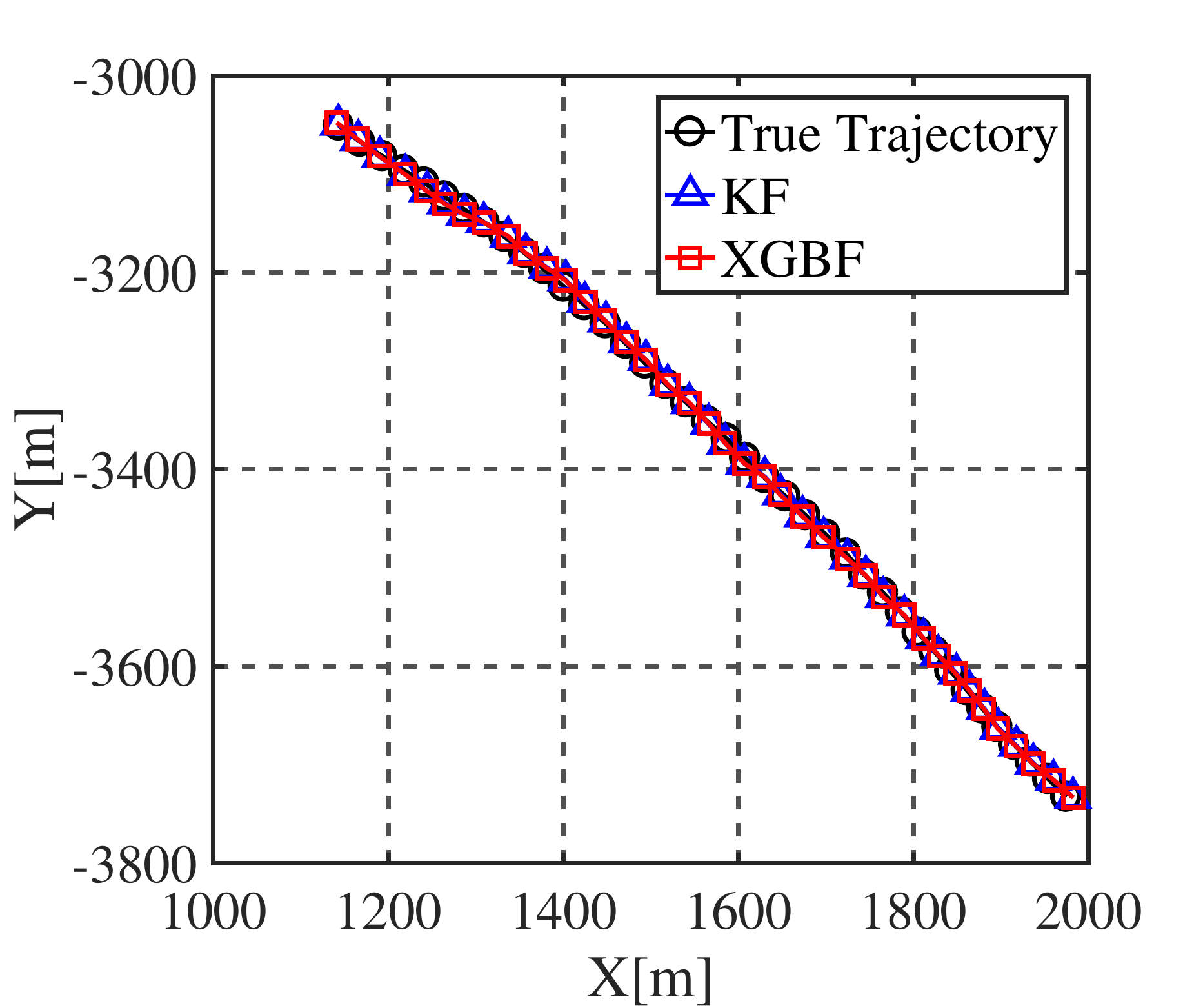}}
\subfigure[]{\includegraphics[width=0.8\columnwidth,draft=false]{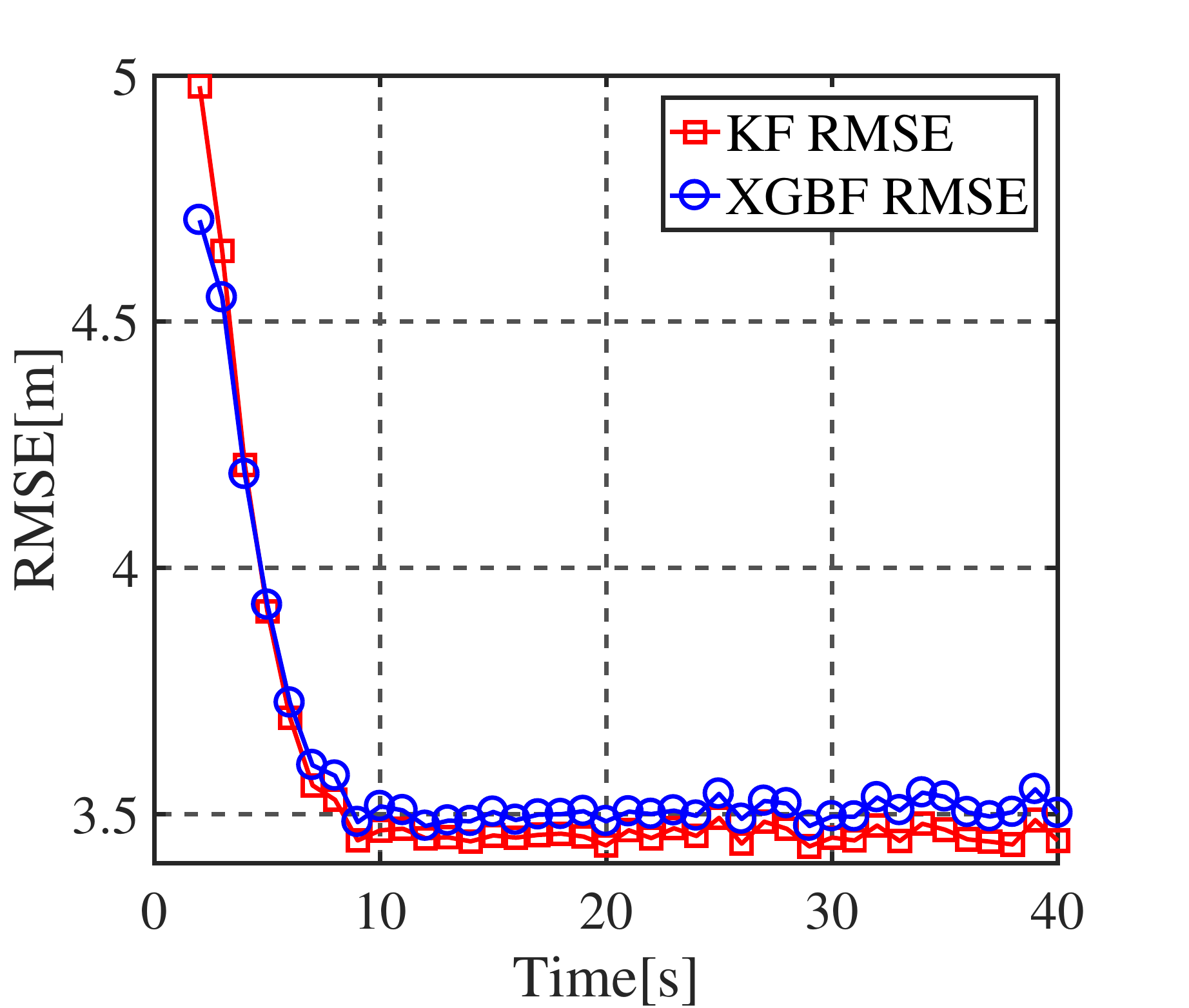}}
\caption{(a) The target track in linear system scenario, reflecting the actual estimated effect of XGBF and KF on the track. (b) The estimated accuracy (RMSE) of KF and XGBF, reflects the value of XGBF based on the linear optimal KF.}
\label{fig:track and RMSE_KF}
\end{figure}

In Fig. \ref{fig:track and RMSE_KF}, the estimation accuracy of XGBF is close to KF in the ``general case''. Because KF is a linear optimal filter, XGBF can approximate the estimation accuracy of KF, which reflects the value of XGBF. In addition, the gap between KF and XGBF will increase slightly during the later stage of filter in Fig. \ref{fig:track and RMSE_KF} (b). This is because the theoretical covariance will gradually decrease when KF changes over time, its RMSE will also converge and become smaller. Thus, the result in Fig. \ref{fig:track and RMSE_KF} (b) is reasonable. However, KF in the real world is unlikely to achieve the theoretical convergence. The detailed simulation is in Section~\ref{simulation:special case}.

\begin{figure}[htbp]
\centering
{\includegraphics[width=0.8\columnwidth,draft=false]{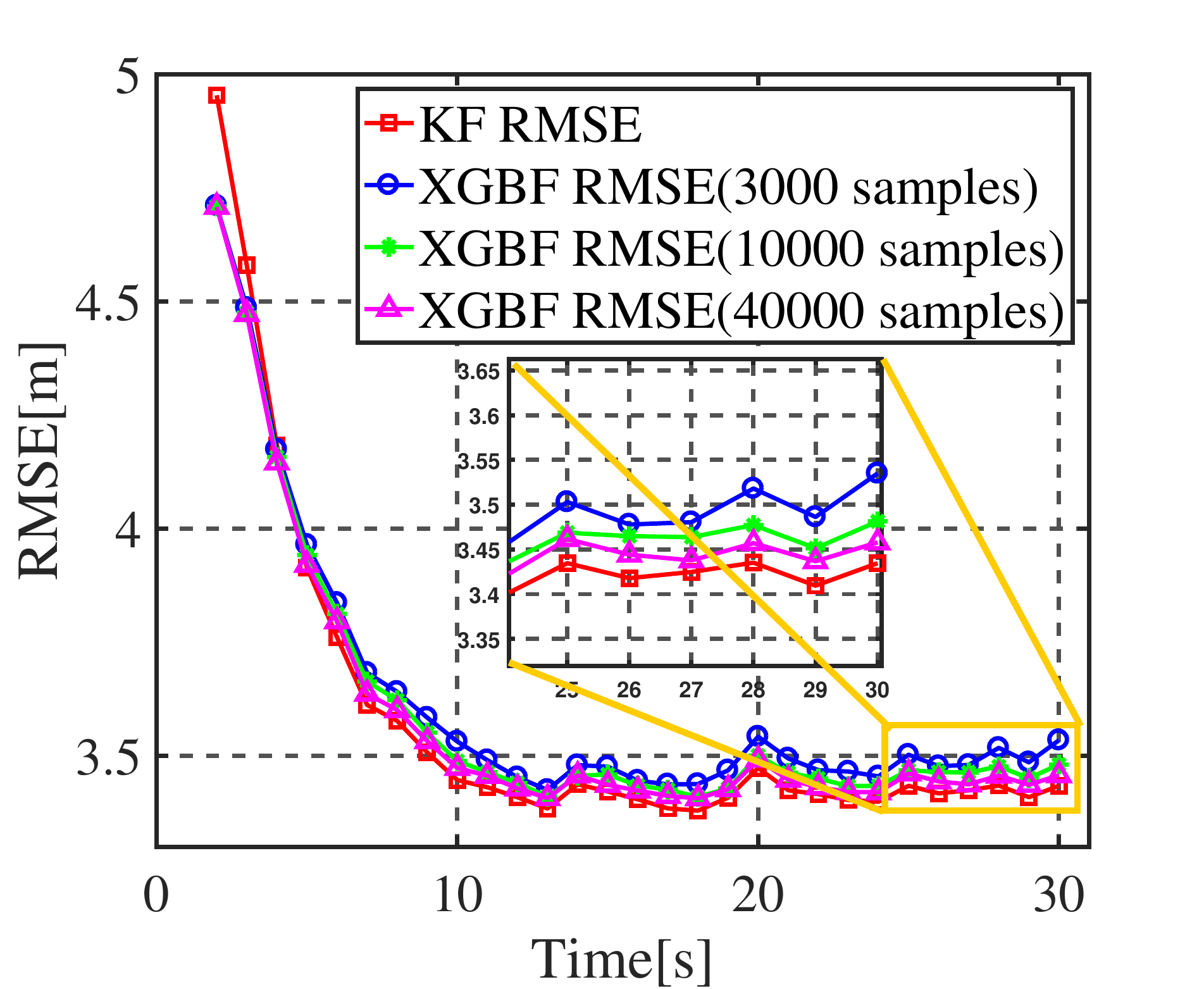}}
\caption{The estimated accuracy (RMSE) of KF and XGBF in larger training samples, which can illustrate the effect of the number of training samples on XGBF.}
\label{fig:sample_KF_XGBF}
\end{figure}

Although it seems that the XGBF's estimation accuracy is slightly lower than KF, this is also due to the lack of training sample. If the training sample is large, the XGBF estimation accuracy will be closer to KF in Fig.\ref{fig:hyperparameters} (b) and Fig. \ref{fig:sample_KF_XGBF}.

In addition, the accuracy improvement is not high in the later stage of the training sample change in Fig. \ref{fig:hyperparameters} (b), which also reflects the effect of our ``sample sparseness confrontation'' method on the number of training samples. More specifically, less training samples in actual scene can also satisfy certain filtering effects.

\subsubsection{KF VS. XGBF in the ``Special Case''}\label{simulation:special case}

In the ``special case'', we consider the following three scenarios:

\begin{itemize}
  \item Scenario $1$: the real scenario is a time-varying process noise environment. In other words, the process noise intensity ${q_s}$ changes continuously with time).
  \item Scenario $2$: the real scenario is the compound noise environment. More specifically, the process noise is an additive composite noise of ``Gaussian $+$ exponent'', the measurement noise is still Gaussian.
  \item Scenario $3$: the real scenario is the compound noise environment. More specifically, the process noise is an additive composite noise of ``Gaussian $+$ exponent'', the measurement noise is a multiplicative noise of ``Gaussian $\times$ exponent''.
\end{itemize}

\begin{figure}[htbp]
\centering
\subfigure[]{\includegraphics[width=0.8\columnwidth,draft=false]{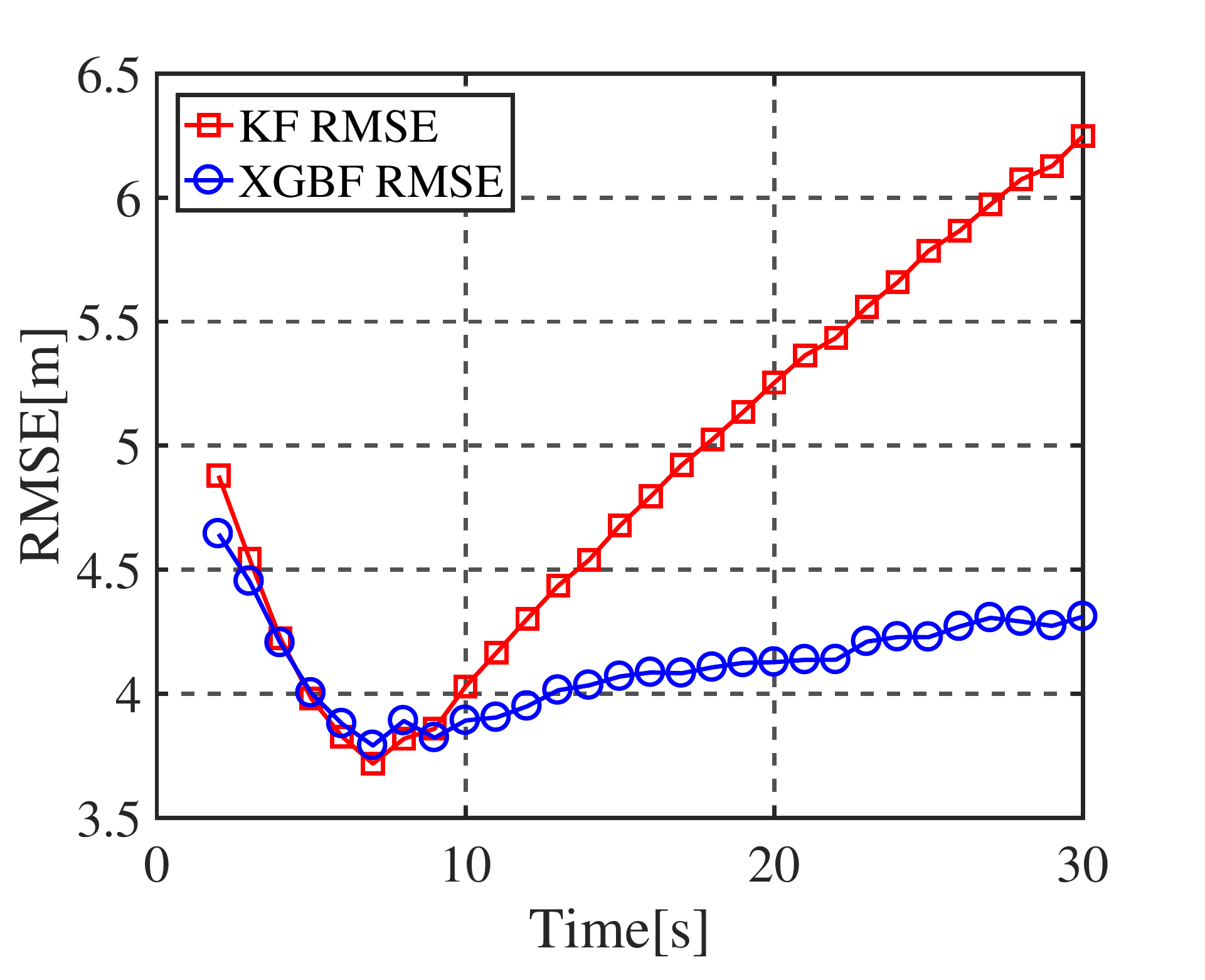}}
\subfigure[]{\includegraphics[width=0.8\columnwidth,draft=false]{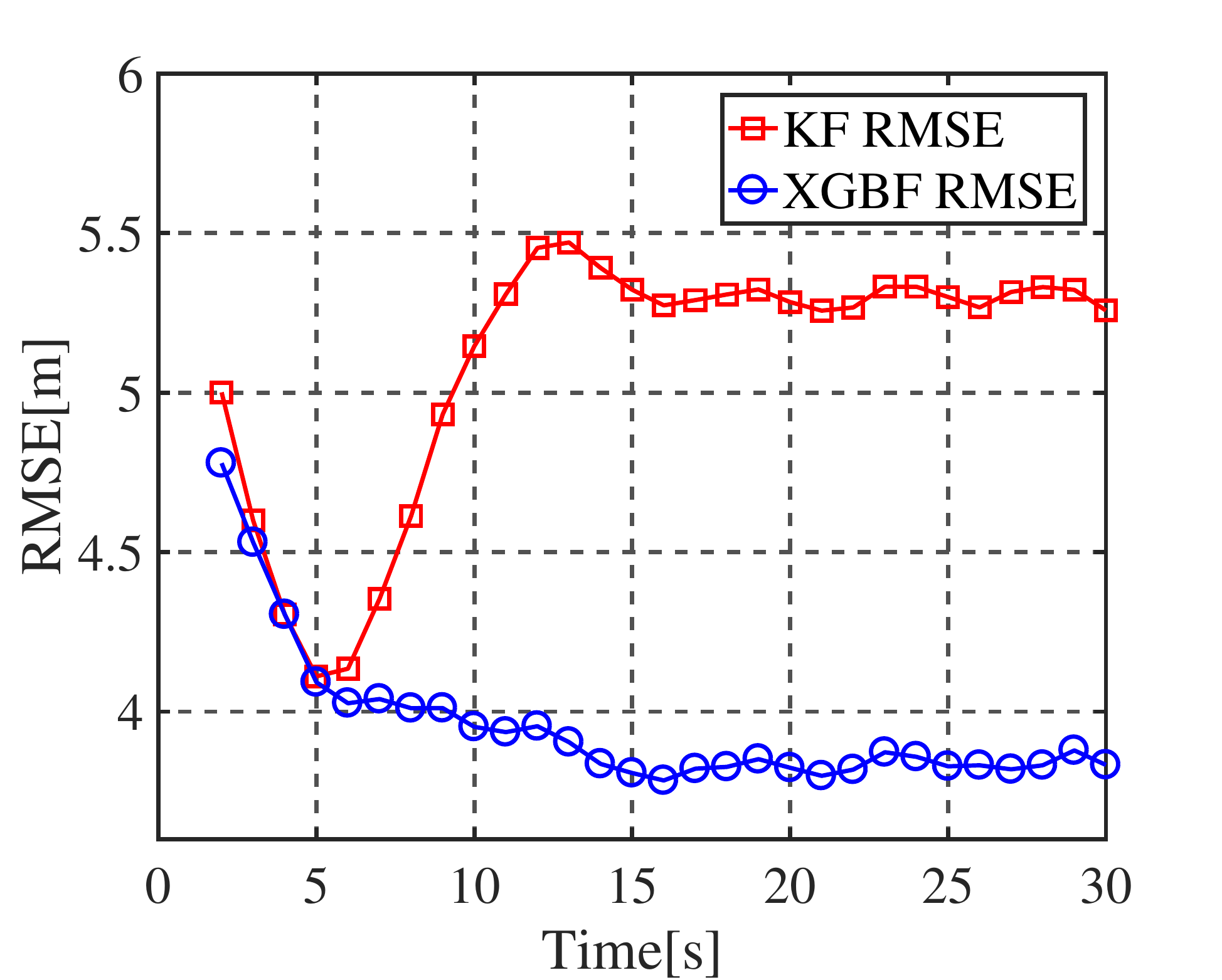}}
\subfigure[]{\includegraphics[width=0.8\columnwidth,draft=false]{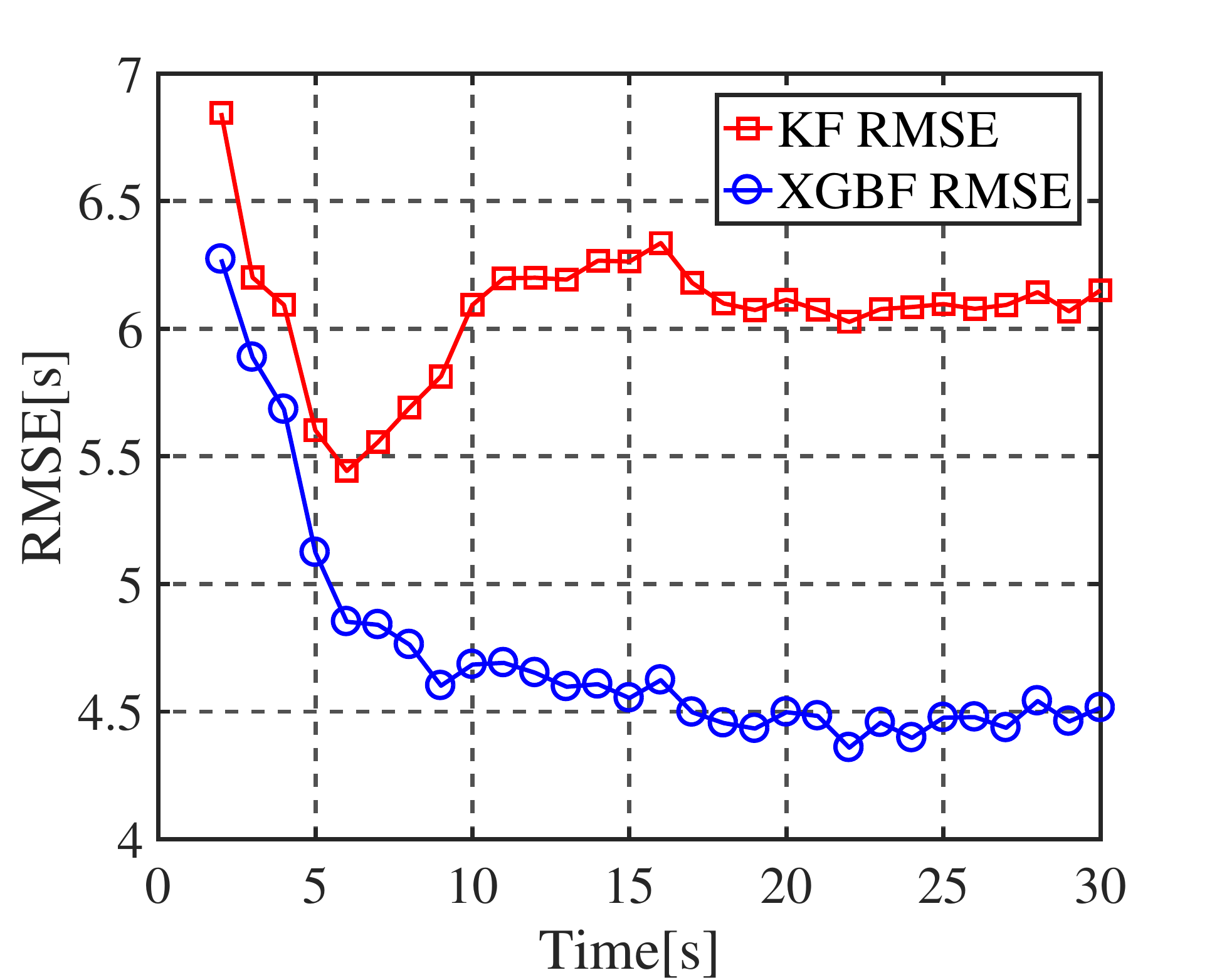}}
\caption{The estimated accuracy (RMSE) of XGBF and KF in the different ``special cases''. (a) Scenario $1$: the real scenario is a time-varying process noise environment. (b) Scenario $2$: the real scenario is the compound noise environment, the process noise is compound noise. (c) Scenario $3$: the real scenario is the compound noise environment, the process noise and the measurement noise are both compound noise.}
\label{fig:special case}
\end{figure}

For scenario $1$, $T$ = 30s, ${w_k} \sim {{\cal N}}\left( {0,{\bf{Q}}} \right)$ and the ${q_s}$ changes with time. ${v_k} \sim {{\cal N}}\left( {0,{\bf{R}}} \right)$ where $v_x^2$ = 30 and $v_y^2$ = 20. For XGBF, the parameters are unchanged. Suppose ${q_s}$ = 0.5 $\times$ $t$, but the KF's model incorrectly estimates ${q_s}$ = 1. Therefore, the estimation accuracy is shown in Fig. \ref{fig:special case} (a).

For scenario $2$, the process noise is an additive composite noise of ``Gaussian $+$ exponent'' as follows
\begin{equation}\label{eq40}
{{\bf{x}}_{k + 1}} = {\bf{F}}{{\bf{x}}_k} + {{\bf{w}}_k} + {{\bf{\cal E}}_k},
\end{equation}
where ${w_k} \sim {{\cal N}}\left( {0,{\bf{Q}}} \right)$ and ${q_s}$ = 1, ${{\bf{\cal E}}_k}$ follows the exponential distribution of the parameter $\kappa $ = 1 (i.e., ${{\bf{\cal E}}_k} \sim {E}\left( {1} \right)$). However, the KF's model incorrectly modeled as only the Gaussian process noise. Therefore, the estimation accuracy is shown in Fig. \ref{fig:special case} (b).

For scenario $3$, the state equation is still \eqref{eq40}, the measurement noise is a multiplicative noise of ``Gaussian $\times$ exponent'' as follows
\begin{equation}\label{eq41}
{{\bf{z}}_k} = {\bf{H}}{{\bf{x}}_k} + {{\bf{v}}_k} \times {{\bf{\cal A}}_k},
\end{equation}
where ${v_k} \sim {{\cal N}}\left( {0,{\bf{R}}} \right)$, ${{\bf{\cal A}}_k} \sim E\left( {1} \right)$. However, the KF's model incorrectly modeled as only the Gaussian process noise and only the Gaussian measurement noise. The simulation result is shown in Fig. \ref{fig:special case} (c).

In Fig.\ref{fig:special case}, we can know that the XGBF of SLF can be well applied to unknown and complicated noise environments. Because XGBF is based on data-driven, which can avoid the problem of the parameter estimation error and the model mismatch. More specifically, when the specific target motion model cannot be determined, the XGBF of SLF uses the data to establish a ``hidden model'' to find the internal mapping relationship and to achieve good estimation accuracy. However, the traditional KF has a greatly reduced accuracy due to depend heavily on the prior information of the model. Therefore, this special case accurately reflects the value and advantages of XGBF.

\subsubsection{The Effects of ${\bf{Q}}$ and ${\bf{R}}$}

Suppose $T$ = 30s. For XGBF, the number of training and test samples and other parameters are unchanged.

\begin{figure}[htbp]
\centering
\subfigure[]{\includegraphics[width=0.8\columnwidth,draft=false]{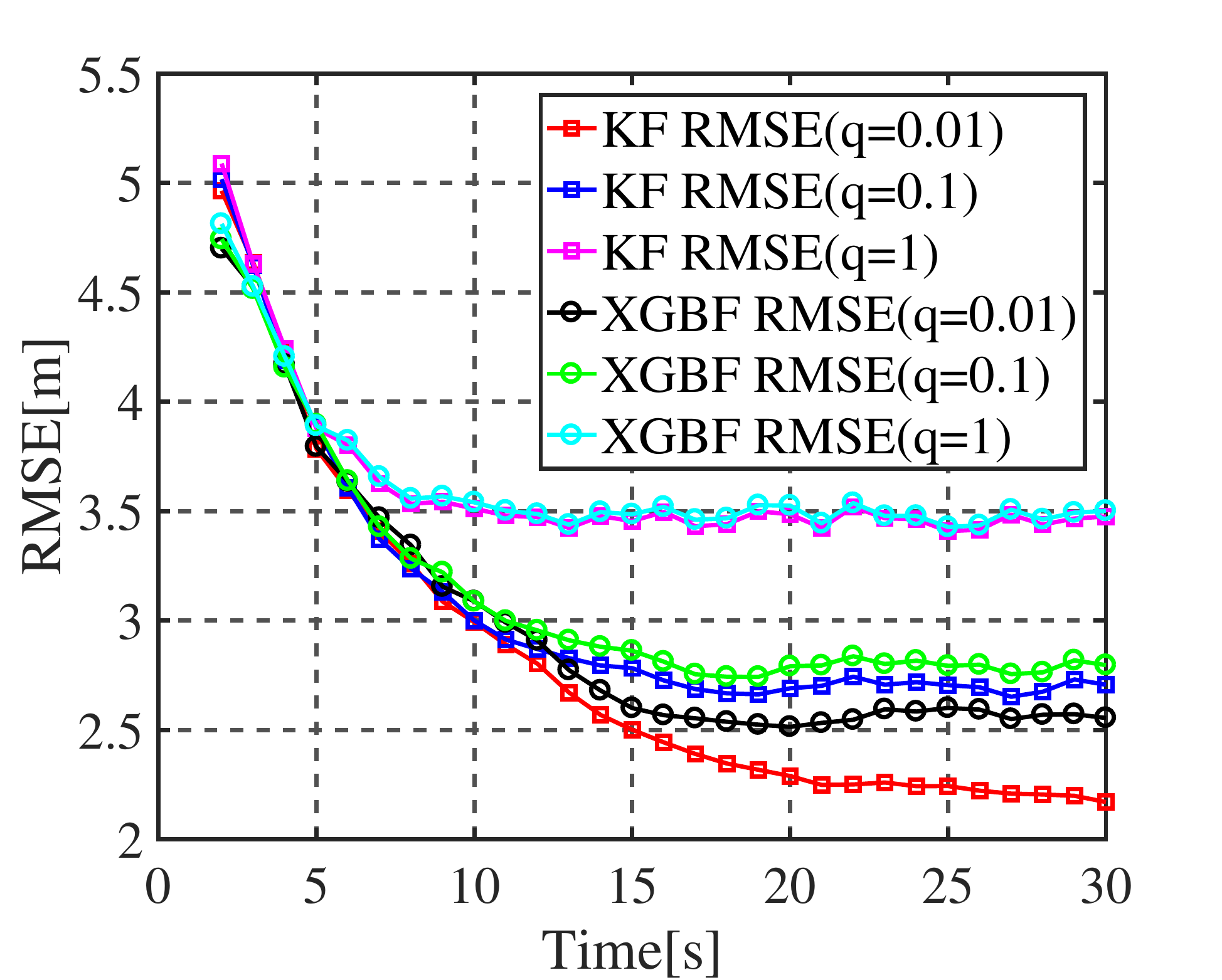}}
\subfigure[]{\includegraphics[width=0.8\columnwidth,draft=false]{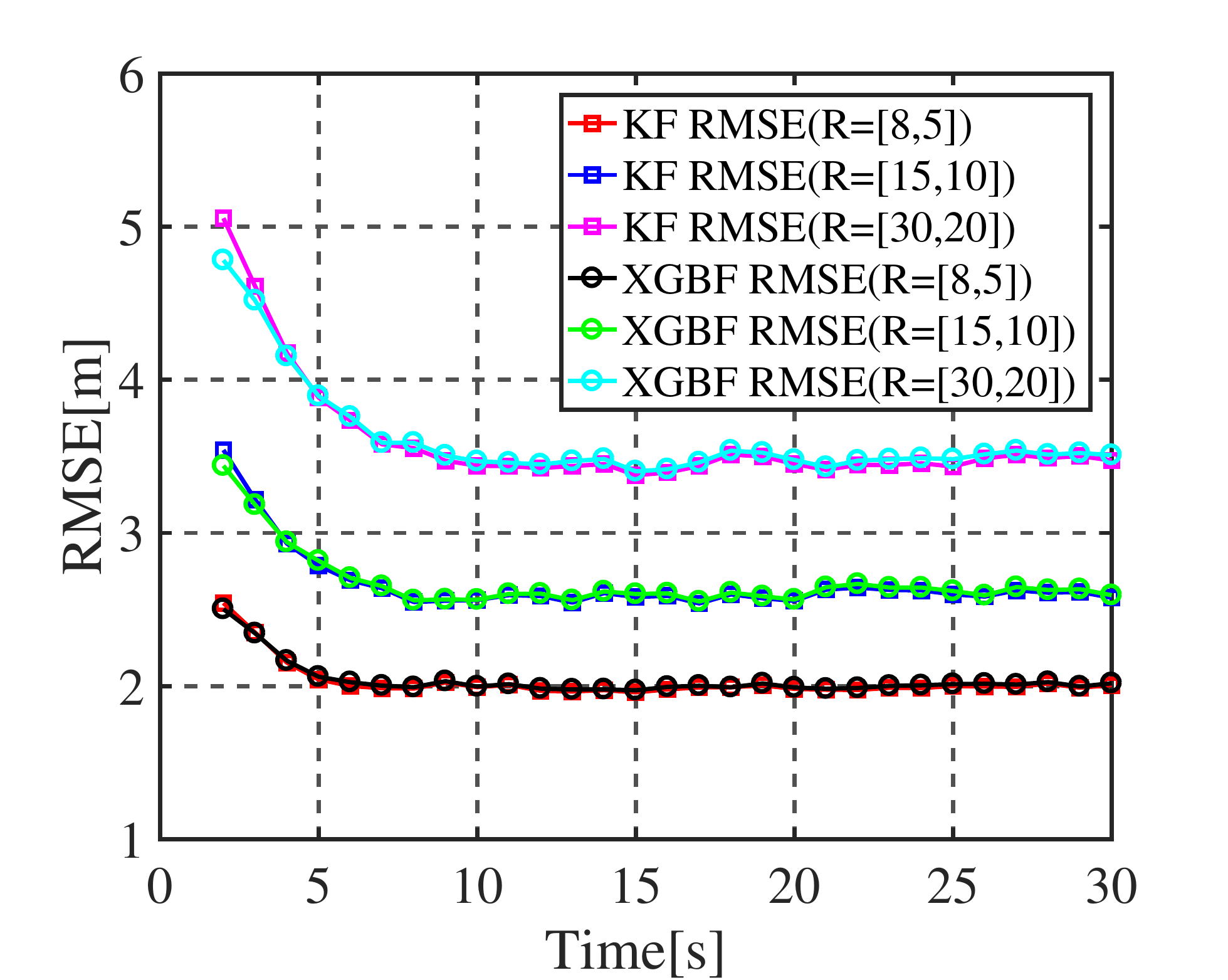}}
\caption{The effects of different noises on XGBF and KF. (a) The effect of ${\bf{Q}}$ on XGBF and KF. (b) The effect of ${\bf{R}}$ on XGBF and KF.}
\label{fig:qR_difference}
\end{figure}

When researching the effect of ${\bf{Q}}$, set ${q_s}$ = 0.01, ${q_s}$ = 0.1, ${q_s}$ = 1 and ${q_s} $ = 3 in the process noise. For the measurement noise, ${\bf{R}}$ = diag\{30, 20\}. The estimation accuracy is shown in Fig. \ref{fig:qR_difference} (a).

When researching the effect of ${\bf{R}}$, set ${\bf{R}}$ = diag\{3, 2\}, ${\bf{R}}$ = diag\{8, 5\}, ${\bf{R}}$ = diag\{15, 10\}, ${\bf{R}}$ = diag\{30, 20\} in the measurement noise. For the process noise, ${q_s}$ = 1. The estimation accuracy is shown in Fig. \ref{fig:qR_difference} (b).

In Fig. \ref{fig:qR_difference}, we can know that the larger the noise and the lower the XGBF estimation accuracy. This is because the larger the noise, the greater the maneuverability of generated track and the larger the measurement error. XGBF is based on the ``data driven'' and only considers the mapping relationships within the data. Therefore, KF will have better estimation accuracy when the process noise is very small.

\section{Conclusions}\label{Sec: conclusion}

In this paper, we give a new solution way for the filter of target tracking using the supervised learning idea, which do not need to build the motion model and evaluate model parameters. More specifically, we construct a supervised learning based online tracking filter (SLF) framework, laying a foundation for the further application of supervised learning to traditional tracking problems.

The key of our approach is to consider the sample sparseness problem, which is convenient for training and learning from the target motion information. Next, establish a ``hidden model'' based on the training data to find the mapping relationship within the data. Thus, SLF has the ability to avoid the modeling mismatch problem that may lead to a performance decrease in traditional tracking method. Then, we use XGBoost as the specific implementation method of SLF, which propose XGBF. Finally, simulation experiments show that the proposed XGBF
still has good estimation accuracy compared to KF in the complicated unknown noise environments. Meanwhile, the experiments show the effectiveness and robustness of our approach.

\bibliographystyle{IEEEtran}
\bibliography{ref}

\end{document}